\documentclass[12pt,times,a4paper,floatfix]{iopart}
\usepackage{iopams}
\usepackage{amsfonts}
\usepackage{tikz}
\usepackage{epstopdf}
\usepackage{graphicx}
\usepackage[caption=false]{subfig}
\usepackage{pgf,tikz}
\usepackage{float}
\usepackage{color}
\usepackage{hyperref}
\usepackage{lineno}
\usepackage{cite}
\usetikzlibrary{arrows}
  
\begin{document}
\title[]{Relativistic sonic geometry for isothermal accretion in the Kerr metric}
\author{Md Arif Shaikh}

\address{Harish-Chandra Research Institute, HBNI, Chhatnag Road, Jhunsi, Allahabad 211019, India}
\ead{\mailto{arifshaikh@hri.res.in}}

\begin{abstract}
We linearly perturb advective isothermal transonic accretion onto rotating astrophysical 
black holes to study the emergence of the relativistic acoustic spacetime and to 
investigate how the salient features of such spacetime get influenced by the spin 
angular momentum of the black hole. We have perturbed three different quantities - the velocity potential, the mass accretion rate and the relativistic Bernoulli's constant to show that the acoustic metric obtained for these three cases are same up to a conformal factor . By constructing the required causal structures, it has been demonstrated that 
the acoustic black holes are formed at the transonic points of the flow and acoustic white holes 
are formed at the shock location. The corresponding acoustic surface gravity has been computed in terms 
of the relevant accretion variables and the background metric elements. The linear stability 
analysis of the background stationary flow has been performed.
\end{abstract}


\vspace{2pc}

\submitto{\CQG}

\section{Introduction}
Classical analogue systems are those where a curved acoustic geometry may be obtained by linear perturbing a transonic fluid. Such sonic geometries are embedded within the background fluid, and the propagation of the acoustic perturbation can be described by an acoustic metric which possesses sonic horizons \cite{Unruh,Visser1998,Bilic1999,Barcelo,Analogue-gravity-phenomenology,Unruh-schutzhold,Novello-visser}. Accreting astrophysical black holes provide an embedding for an acoustic spacetime, which is embedded within the fluid system accreting on the black hole, and thus such systems are very special examples of classical analogue gravity systems, since such configuration may be considered unique because both the gravitational as well as acoustic horizons are an integral part of such systems \cite{Das-2004,abraham:causal,Dasgupta-bilic-das-2005,Das-bilic-dasgupta,Pu-maity-das-chang,Bilic-Choudhury,Tarafdar-das-2015,arif:iso_sch,deepika-sph2015,deepika_ax_schwarzchild,Satadal-arif-2016}. In the majority of such works, characteristic features of the embedded sonic geometry have been obtained through perturbation of mass accretion rate in the astrophysical context. 

Shaikh \etal \cite{arif:iso_sch} have studied the perturbation of isothermal accretion in a more general context, where a generalized formalism was presented to show how one can obtain the corresponding sonic geometry through the linear perturbation of the velocity potential, mass accretion rate as well as the relativistic Bernoulli's  constant to manifest that there exist some general properties of the emergent gravity phenomena which are independent of the quantity that we linear perturb to obtain the spacetime geometry describing the emergence of the gravity like phenomena. In \cite{arif:iso_sch}, the accretion dynamics was studied in the background Schwarzschild metric. Astronomers, however, believe that most of the astrophysical black holes possess non zero spin angular momentum, i.e, such black holes will be characterized by non zero Kerr parameter `$a$' (e.g., see \cite{Miller2009,Kato2010,Ziolkowski2010,Tchekhovskoy2010,
Daly2011,Buliga2011,Reynolds2012,McClintock2011,Martínez-Sansigre2011,Dauser2010,Nixon2011,McKinney2012,
McKinney2013,Brenneman,Rolando2013,Sesana,Fabian2014,
Healy,Jiang,Nemmen} which studies the influence of black hole spin and measurement of such spin for different astrophysical systems). 

We would thus like to understand how the emergent gravity phenomena and various characteristic features of the sonic geometry get influenced by the black hole spin. In other words, in the present work, we would like to demonstrate how the property (spin) of the background metric influences the properties of the emergent acoustic metric. Saha \etal \cite{Sonali2016} have studied the dependence of few of the properties of the acoustic metric on the spin parameter, however, such studies were done for accretion under the influence of generalized post-Newtonian pseudo-Kerr black hole potential and the studies in the case of isothermal flow was very limited. Whereas our work is in fully general relativistic Kerr background and concentrated on isothermal flow. Important discussion on shocked isothermal accretion in the Kerr geometry could be found in \cite{Yuan1996} but the corresponding acoustic geometry was not discussed in \cite{Yuan1996} which we study in the present work.  Also, they discuss only the stationary solution, whereas here we study the time dependence through linear stability analysis, which has not been done before. In \cite{TKD2012-Czerny}, it was shown that stationary shock formation is an integral part of accretion disc in the Kerr metric and in the present work we show how shock formation influences the properties of the acoustic spacetime, e.g., the causal structure of acoustic spacetime. We assume an inviscid flow as viscosity breaks the Lorentzian symmetry \cite{Barcelo} and acoustic metric cannot be constructed. However, for relativistic viscous transonic accretion flow onto rotating black hole interested readers are  referred to  \cite{Peitz1997,Takahashi}. 

Accretion of the surrounding matter onto astrophysical black holes is a crucial phenomenon to study in connection with the observational evidence of the black holes in the universe. While falling onto the black hole, accreting matter emits multi-wavelength radiation due to various radiative processes. The details of the characteristic features of such emitted photons may be understood by studying various dynamical properties of the accretion flow as well as by applying the knowledge of the radiative transfer. Such emitted photons form its characteristic spectra. Study of such spectra provides the information about the strong gravity spacetime close to the black hole event horizon, and, most importantly, helps ensure the presence of astrophysical black holes through observational means. Apart from the wind-fed accretion (where the supersonic stellar wind accretes onto black holes), the infalling material usually starts its journey toward the black hole subsonically. Close to the event horizon of the black hole, the dynamical velocity becomes extremely large, close to the speed of light in the vacuum ($ c $). Whereas, even for the steepest equation of state, the maximally allowed velocity of sound within the accreting matter never exceeds $ \frac{c}{\sqrt{3}} $. This dictates that the accreting matter falls supersonically onto the black hole.

The black hole accretion is, thus, necessarily transonic, and hence matter makes a smooth (continuous) transition from subsonic to supersonic state. Such transonic property of the accreting material thus suggests that an acoustic geometry with the (at least one) acoustic horizon may be present within the accreting material, where the transonic surface may act as the corresponding acoustic horizon specified by the acoustic metric.

If the accretion is spherically symmetric, then the formation of more than one sonic horizon is not possible. For axially symmetric accretion flow, however, one may argue that two saddle type critical points may form and the accreting flow may make a smooth continuous transition from a subsonic to a supersonic state and hence two acoustic black hole horizons may form. In between these two sonic surfaces, a stationary shock forms which forces the flow to make a discontinuous transition from a supersonic state to a subsonic state. Such shock surfaces may be identified with an acoustic white hole. Shock transition is considered to be a crucial phenomenon in accretion astrophysics. The discontinuous changes of the values of various variables may affect the emergent spectra by introducing (possibly) a broken power law since the photon energy changes abruptly. The shock formation phenomena in black hole accretion disc thus explicitly manifest through the observation of the characteristic black hole spectra.

It is thus obvious that the accreting black hole system has a one to one correspondence with a classical analogue system naturally found in the universe. Hence it is important to study the accreting black hole system from the analogue gravity point of view. Usually, a steady state accretion flow is considered to understand various spectral features of the astrophysical black holes. It is imperative to ensure that such a task can be accomplished through the linear perturbation technique. The stationary integral solutions of the steady state accretion flow are linear perturbed to demonstrate that such perturbation does not diverge and hence the steady state remains stable. 

In analogue gravity phenomena, a stationary state is linear perturbed and the propagation of the resultant acoustic perturbation is described by the acoustic metric. For astrophysical accretion, one has to show that the linear perturbation of the stationary accretion solutions leads to the emergence of the acoustic metric embedded within the accreting matter. Such acoustic metric will have acoustic horizons. By constructing the causal structures, one has to show that such horizons actually coincide with the physical transonic surfaces in the accretion flow, and the shock surface coincides with the acoustic white whole horizon. In this way, a formal correspondence can be made between an accreting astrophysical black hole and a classical analogue gravity system. Such task has been accomplished in the present work for general relativistic, axially symmetric, isothermal inviscid accretion onto rotating (Kerr type) astrophysical black holes. 

We will consider general relativistic sub-Keplerian low angular momentum inviscid accretion of isothermal matter onto a Kerr black hole as the background steady flow. Such flow is considered to be characterized by the constant bulk temperature $T$ (ion temperature of single temperature flow), constant specific angular momentum $\lambda$, and the Kerr parameter $a$. We consider the low angular momentum accretion to fulfill the inviscid ( and hence constant $\lambda$) criteria as flow with large angular momentum may settle into a Keplerian orbit and in absence of viscous transport of angular momentum the accretion may not happen.  However, an inviscid flow may not necessarily be sub-Keplerian as there may be other mechanisms (non-viscous) which may drive the accretion process. In the present accretion model, we do not include these mechanisms and hence restrict ourselves to low angular momentum flow.  A large pool of works is available in the literature which deals with effectively inviscid flow \cite{Abramowicz1981,JFLU1985,Blaes1987,Fukue1987,Anderson1989,Beloborodov,Kafatos1994,Yang1995,Pariev1996,JFLu1997,Okuda2007,Nagakura2008,Nagakura2009,Janiuk2009,Sukova2015MNRAS,Sukova2015JOP,Sukova2017}. Also, see \cite{Abramowicz2013} for a general review of accretion disk.  Such practically inviscid flow may be observed at our galactic center \cite{Monika}. Hence the assumption of inviscid accretion is not unjustified from astrophysical context. The specific angular momentum of the axially symmetric rotating flow and the spin angular momentum (the Kerr parameter) of the astrophysical black holes are two effectively additive quantities in our work. This means, if one reduces the Kerr parameter (the algebraic value of the Kerr parameter), then the effective value of the maximum angular momentum $ \lambda $ for which a multi-transonic flow will form, will increase, and vice-versa. This is probably due to the fact that the frame-dragging effect does not explicitly influence our calculations, and hence the parameter $ a $ effectively `adds up' with $ \lambda $. 

The plan of the work is as follows. We shall first construct the general relativistic Euler and continuity equation corresponding to the ideal fluid described by the isothermal equation of state. We shall find the integral solutions of the time-independent part of the Euler and the continuity equation to obtain two first integrals of motion for background steady flow. Such constants of motion are found to be the mass accretion rate $\Psi_0$ obtained as the integral solution of the continuity equation and the relativistic Bernoulli's constant $\xi_0$ which is obtained through the integral solution of the Euler equation. We then introduce the irrotationality condition which provides the constant steady value of the velocity potential $\psi$. We then linear perturb $\psi$, $\Psi$ and $\xi$ separately and find that the overall characteristics features of the embedded sonic geometries are more or less same. They might differ in the conformal factor (which may appear as the coefficient of the acoustic metric element) only.

Once the acoustic metric is found, the location of the acoustic horizon is identified. Considering the detailed phase portrait of the characteristic stationary integral solutions, we show that at most three acoustic horizons may be present in such configurations. Out of those three horizons, two are black hole horizons formed at the outer and the inner transonic points and one also obtains an acoustic white hole which is obtained at the location where a discontinuous stationary shock transition takes place. We then construct the causal structures at and around such acoustic horizons. 
We demonstrate how to calculate the acoustic surface gravity in terms of the background black hole metric elements and corresponding accretion variables, all computed at the location of the corresponding acoustic horizon on which the value of the acoustic surface gravity is being evaluated.

Finally, we perform the stability analysis of the stationary solutions used for the calculation of the causal structure. Such analysis ensures that the stationary solutions that we are using are stable under linear perturbation.

We shall set $G=c=M_{\rm BH}=1$ where $G$ is the universal gravitational constant, $c$ is the velocity of light and $M_{\rm BH}$ is the mass of the black hole. The radial distance will be scaled by $G M_{\rm BH}/c^2$ and any velocity will be scaled by $c$. We shall use the negative-time-positive-space metric convention.

\section{Basic equations}
\subsection{The background metric}
We Consider the following metric for a stationary rotating space time
\begin{equation}\label{background_metirc}
ds^2=-g_{tt}dt^2+g_{rr}dr^2+g_{\theta\theta}d\theta^2+2g_{\phi t}d\phi dt+g_{\phi\phi}d\phi^2,
\end{equation}
where the metric elements are function of $r,\theta$ and $\phi$. The metric elements in the Boyer-Lindquist coordinates are given by \cite{Bardeen1972,Poisson2004relativist}
\begin{equation}\label{back_metric_elements}
\fl g_{tt}=(1-\frac{2r}{\mathcal{A}}),\quad
 g_{rr}=\frac{\mathcal{A}}{\mathcal{B}},\quad
 g_{\theta\theta}=\mathcal{A},\quad
 g_{\phi t}=-\frac{2ar\sin^2\theta}{\mathcal{A}},\quad
 g_{\phi\phi}=\frac{\mathcal{C}}{\mathcal{A}}\sin^2\theta
\end{equation}
where
\begin{equation}\label{A-B-C}
\mathcal{A}=r^2+a^2\cos^2\theta,\quad \mathcal{B}=r^2-2r+a^2,\quad \mathcal{C}=(r^2+a^2)^2-a^2\mathcal{B}\sin^2\theta.
\end{equation}
The event horizon of the Kerr black hole is located at $ r_+ = 1+\sqrt{1-a^2}  $.

\subsection{Conservation equations}
The energy momentum tensor for perfect fluid is given by
\begin{equation}\label{Energy_momentum_tensor}
 T^{\mu\nu}=(p+\varepsilon)v^\mu v^\nu+pg^{\mu\nu}
 \end{equation}
 where $p$ and $\varepsilon$ are the pressure and the mass-energy density of the fluid respectively. $v^\mu$ is the four-velocity of the fluid. $v^\mu$ obeys the normalization condition $v^\mu v_\mu=-1$. We assume the fluid to be inviscid and irrotational and is described by the isothermal equation of state, i.e., $ p \propto \rho $, where $\rho$ is the local rest mass energy density of the fluid. The mass-energy density $ \varepsilon $ includes the rest mass density and the internal energy (thermal energy), 
$ \varepsilon = \rho+\varepsilon_{\rm thermal} $. The rest mass energy density $ \rho $ could be related to the particle number density of fluid $ n $ by the relation $ \rho = m_0 n $, where $ m_0 $ is the rest mass of the fluid particle. The continuity equation which ensures the conservation of mass is given by
\begin{equation}\label{Continuity_eq}
\nabla_\mu(\rho v^\mu)=0
\end{equation}
where the covariant divergence is defined as $\nabla_\mu v^\nu = \partial_\mu v^\nu + \Gamma^\nu_{\mu\lambda}v^\lambda$. The energy-momentum conservation equation is given by
\begin{equation}\label{mom_consv_eq}
\nabla_\mu T^{\mu\nu}=0.
\end{equation}
Substitution of Eq. (\ref{Energy_momentum_tensor}) in Eq. (\ref{mom_consv_eq}) provides
 the general relativistic Euler equation for barotropic ideal fluid as
\begin{equation}\label{Euler_eq}
(p+\varepsilon)v^\mu \nabla_\mu v^\nu+(g^{\mu\nu}+v^\mu v^\nu) \nabla_\mu p=0.
\end{equation}
The enthalpy is defined as
\begin{equation}\label{enthalpy}
h=\frac{(p+\varepsilon)}{\rho}.
\end{equation}
The isothermal sound speed could be defined as\cite{Yuan1996}
\begin{equation}\label{sound_speed}
c_s^2 = \frac{1}{h}\frac{\partial p}{\partial \rho}.
\end{equation}
The relativistic Euler equation for isothermal fluid can thus be written as
\begin{equation}\label{Euler_eq_for_isothermal}
v^\mu \nabla_\mu v^\nu+\frac{c_s^2}{\rho}(v^\mu v^\nu+g^{\mu\nu})\partial_\mu \rho=0.
\end{equation}
For general relativistic irrotational fluid we have the irrotationality condition given by \cite{arif:iso_sch}
\begin{equation}\label{irrotationality_condition}
\partial_\mu(v_\nu\rho^{c_s^2})-\partial_\nu(v_\mu\rho^{c_s^2})=0.
\end{equation}
\section{The accretion disc structure and the reference frame}
\subsection{Disc structure}
The accretion flow is described by the four-velocity $(v^t,v^r,v^\theta,v^\phi$). We assume the flow to be axially symmetric and also to be symmetric about the equatorial plane and the velocity along the vertical direction to be negligible, i.e., $v^\theta\approx 0$. Now let us consider the continuity equation given by (\ref{Continuity_eq}). Due to the axial symmetry, the $\frac{\partial}{\partial \phi}$ term would vanish. It is a common practice for accretion flow with low angular momentum to do a vertical averaging of the flow \cite{gammie_and_popham}. The vertical averaging of a flow variable $f(r,\theta)$ is approximated as
\begin{equation}\label{vertical-averaging}
\int f(r,\theta)d\theta \approx H_\theta f(r,\theta=\frac{\pi}{2})
\end{equation}
where $H_\theta$ is the characteristic angular scale of the flow. Thus the continuity equation for vertically averaged axially symmetric accretion can be written as\cite{gammie_and_popham,deepika_ax_schwarzchild,arif:iso_sch}
\begin{equation}\label{continuity_eq_equatorial}
 \partial_t(\rho v^t \sqrt{-\tilde{g}}H_\theta)+\partial_r(\rho v^r \sqrt{-\tilde{g}} H_\theta)=0
\end{equation}
where $\tilde{g}$ is the value of $g$, the determinant of the background metric $g_{\mu\nu}$, on the equatorial plane. For Kerr metric $g=-\sin^2\theta\mathcal{A}^2$ and thus $\tilde{g}=-r^4$. The term $H_\theta$, as mentioned above, appears as a result of the vertical averaging. Vertical averaging of flow allows us to work fully in the equatorial plane by retaining the
information of the vertical structure in the  $H_\theta$ term. In other words, $H_\theta$ is the proper weight function to give correct value of the mass accretion rate obtained by spatially integrating Eq. (\ref{continuity_eq_equatorial}) in stationary case. Thus $H_\theta$ can be related to the local flow thickness $H(r)$ of the flow. In the present work, we consider the flow to be wedge-shaped conical flow where $H(r)\propto r$, where $r$ is the local radius. For a flow with conical geometry thus $H_\theta\propto \frac{H}{r}=\rm{constant}$. $H_\theta$
does not depend on the accretion variables like density and velocities. Thus linear perturbation of these quantities (as defined by the Eq. (\ref{pert_v}),(\ref{pert_rho}) (to be searched below)) will have no effect on it. For simplicity, therefore, we will write $H_\theta$ simply as $H_0$. Further details on different flow geometries can be found in \cite{Bilic-Choudhury,deepika_ax_schwarzchild,arif:iso_sch}. From now on all the equations will be derived by assuming the flow to be vertically averaged and the variables have values equal to that in the equatorial plane. 

\subsection{Reference frame}
Apart from the Boyer-Lindquist coordinate frame (BLF) we shall use a second reference frame which is called the corotating frame (CRF)\cite{gammie_and_popham}. This frame is obtained by an azimuthal Lorentz boost from the locally nonrotating frame (LNRF) into a tetrad basis that corotates with the fluid. LNRF is an orthonormal tetrad basis who lives at $ \theta={\rm constant}, r= {\rm constant}, \phi=\omega t + {\rm constant} $, where $ \omega = \frac{2a}{r^3+a^2(r+2)} $ on the equatorial plane (originally calculated by Bardeen \etal  \cite{Bardeen1972}). Let $ u $ be the radial velocity (referred as the `advective velocity') of the fluid as measured 	in the CRF and $ \lambda = -\frac{v_\phi}{v_t} $ be the specific angular momentum of the fluid. Then the four-velocity components in BLF is related to $ u $ and $ \lambda $ in the following way \cite{gammie_and_popham}

\begin{equation}\label{v_in_CRF}
v^r=\frac{u}{\sqrt{g_{rr}(1-u^2)}}
\end{equation}

\begin{equation}\label{vt_0}
v^t=\sqrt{\frac{(g_{\phi\phi}+\lambda g_{\phi t})^2}{(g_{\phi\phi}+2\lambda g_{\phi t}-\lambda^2g_{tt})(g_{\phi \phi}g_{tt}+g_{\phi t}^2)(1-u^2)}}
\end{equation}
and 
\begin{equation}\label{v_t}
v_t = -\sqrt{\frac{g_{tt}g_{\phi\phi}+g_{\phi t}^2}{(g_{\phi\phi}+2\lambda g_{\phi t}-\lambda^2 g_{tt})(1-u^2)}}
\end{equation}
$ u $ and $ \lambda $ are the two velocity variables needed to describe the flow in the equatorial plane. 
\section{Linear perturbation of velocity potential, mass accretion rate and the relativistic Bernoulli's constant}\label{Sec:Linear_pert}
In the following sections, we introduce three quantities that we want to linear perturb to obtain the acoustic metric. These quantities are - the velocity potential which is defined through the irrotationality condition, the mass accretion rate which is the integral of the continuity equation for stationary flow and the relativistic Bernoulli's constant which is the integral of the temporal component of the relativistic Euler equation for stationary flow introduced in Sec. \ref{sec:velo},\ref{sec:mass} and \ref{sec:bernoulli} respectively. After that, we use the perturbation equations given by Eq. (\ref{pert_v})-(\ref{pert_rho}) in the continuity equation, irrotationality condition and the temporal component of the relativistic Euler equation and keep only the terms that are linear in the perturbed quantities. These equations give the wave equations describing the propagation of the perturbation of the three quantities mentioned above. From these wave equations, we identify the matrix $f^{\mu\nu}$, as discussed later, to obtain the acoustic metric. Thus the main aim of this section is to find the expressions for the symmetric matrix $f^{\mu\nu}$ for three different quantities.

Before going to the details of the derivation of the acoustic metric by linear perturbation of different quantities, let us first write down some useful equations which will be essential in the following sections.
The normalization condition $v^\mu v_\mu=-1$ gives
\begin{equation}\label{Normalization_condition}
g_{tt}(v^t)^2=1+g_{rr}(v^r)^2+g_{\phi\phi}(v^\phi)^2+2g_{\phi t}v^\phi v^t
\end{equation}
From irrotaionality condition given by Eq. (\ref{irrotationality_condition}) with $\mu=t$ and $\nu=\phi$ and with axial symmetry we have
\begin{equation}\label{irrotaionality_t_phi}
\partial_t(v_\phi\rho^{c_s^2})=0
\end{equation}
again with $\mu=r$ and $\nu=\phi$ and the axial symmetry the irrotationality condition gives
\begin{equation}\label{irrotationality_r_phi}
\partial_r(v_\phi\rho^{c_s^2})=0
\end{equation}
So we get that $v_\phi \rho^{c_s^2}$ is a constant of the motion. Eq. (\ref{irrotaionality_t_phi}) gives
\begin{equation}\label{del_t_v_phi}
\partial_t v_\phi=-\frac{v_\phi c_s^2}{\rho}\partial_t \rho
\end{equation}
Substituting  $v_\phi=g_{\phi\phi}v^\phi+g_{\phi t}v^t$ in the above equation provides
\begin{equation}\label{del_t_v_up_phi}
\partial_t v^\phi=-\frac{g_{\phi t}}{g_{\phi\phi}}\partial_t v^t-\frac{v_\phi c_s^2}{g_{\phi\phi}\rho}\partial_t \rho
\end{equation}
Also differentiating the Eq. (\ref{Normalization_condition}) with respect to $t$ gives
\begin{equation}\label{del_t_v_up_t_1}
\partial_t v^t=\alpha_1\partial_t v^r+\alpha_2\partial_t v^\phi
\end{equation}
where $\alpha_1=-\frac{g_{rr}v^r}{v_t}$ and $\alpha_2=-\frac{v_\phi}{v_t}$. Substituting $\partial_t v^\phi$ in Eq. (\ref{del_t_v_up_t_1}) using Eq. (\ref{del_t_v_up_phi}) gives \begin{equation}\label{del_t_v_up_t_2}
\partial_t v^t=\left( \frac{-\alpha_2v_\phi c_s^2/(\rho g_{\phi\phi})}{1+\alpha_2 g_{\phi t}/g_{\phi\phi}}\right)\partial_t \rho+\left(\frac{\alpha_1}{1+\alpha_2 g_{\phi t}/g_{\phi\phi}} \right)\partial_t v^r
\end{equation}
We perturb the velocities and density around their steady background values as following
\begin{equation}\label{pert_v}
v^\mu (r,t)=v^\mu_0(r)+v^\mu_1(r,t),\quad \mu=t,r,\phi
\end{equation}
\begin{equation}\label{pert_rho}
\rho(r,t)=\rho_0(r)+\rho_1(r,t)
\end{equation}
where the subscript ``0" denotes the stationary background part and the subscript ``1" denotes the linear perturbations. Using Eq. (\ref{pert_v})-(\ref{pert_rho}) in Eq. (\ref{del_t_v_up_t_2}) and retaining only the terms of first order in perturbed quantities we obtain

\begin{equation}\label{del_t_pert_v_up_t}
\partial_t v_1^t=\eta_1\partial_t \rho_1+\eta_2\partial_t v^r_1
\end{equation}
where
\begin{equation}\label{eta_1_eta_2_and_Lambda}
\eta_1=-\frac{c_{s}^2}{\Lambda v^t_0\rho_0}[\Lambda (v^t_0)^2-1-g_{rr}(v^r_0)^2],\quad \eta_2=\frac{g_{rr}v^r_0}{\Lambda v^t_0}\quad {\rm and}\quad \Lambda=g_{tt}+\frac{g_{\phi t}^2}{g_{\phi\phi}}
\end{equation}

\subsection{Linear perturbation of velocity potential}\label{sec:velo}
The irrotationality condition given by Eq. (\ref{irrotationality_condition}) can be used to introduce a potential field $\psi$ such that
\begin{equation}\label{velocity_potential_def}
-\partial_\mu \psi=v_\mu \rho^{c_s^2}
\end{equation}
It has been shown in the previous section that $v_\phi\rho^{c_s^2}$ is a constant of motion and hence $\partial_\phi \psi$ is a constant of the motion. Therefore perturbation of $\partial_\phi\psi$ gives $\delta(\partial_\phi\psi)=\partial_\phi(\delta\psi)=\partial_\phi \psi_1=0$. Contracting both sides of Eq. (\ref{velocity_potential_def}) with $v^\mu$ and using the normalization condition $v^\mu v_\mu=-1$ gives
\begin{equation}\label{rho_cs_sq}
\rho^{c_s^2}=v^\mu \partial_\mu \psi
\end{equation}
We perform linear perturbation of the above Eq. (\ref{rho_cs_sq}) by substituting $v,v^t,v^\phi,\rho$ using Eq. (\ref{pert_v}),(\ref{pert_rho}) respectively and $\psi$ by the following equation 
\begin{equation}\label{pert_velocity_pot}
\psi=\psi_0+\psi_1
\end{equation} 
where $\psi_0$ is the stationary value of $\psi$ independent of time. Retaining only the terms that are linear in the perturbation quantities gives
\begin{equation}\label{rho_1}
\rho_1=\frac{1}{c_s^2\rho_0^{c_s^2-1}}[v^r\partial_r\psi_1+v^t\partial_t\psi_1]
\end{equation}
In obtaining the above equation we have used $\delta(v^\mu v_\mu)=0$ and $\partial_\phi \psi_1=0$. 
Eq. (\ref{velocity_potential_def}) can also be written as $v^\mu\rho^{c_s^2}=-g^{\mu\nu}\partial_\nu \psi$. 
Writing $v^\mu=v^\mu_0+v^\mu_1$ and using Eq. (\ref{pert_rho}) and Eq. (\ref{pert_velocity_pot}) in the equation $v^\mu\rho^{c_s^2}=-g^{\mu\nu}\partial_\nu \psi$ keeping only terms which are linear in the perturbed terms gives
\begin{equation}
v^{\mu}_1\rho_0^{c_s^2}=-g^{\mu\nu}\partial_\nu \psi_1-v^\mu_0v^\nu_0 \partial_\nu \psi_1
\end{equation}  Therefore we get
\begin{equation}\label{v_t_up_1_velo}
v^t_1=\frac{1}{\rho^{c_s^2}}[(g^{tt}-(v^t_0)^2)\partial_t\psi_1-v^t_0v^r_0\partial_r\psi_1]
\end{equation}
and
\begin{equation}\label{v_1_velo}
v^r_1=\frac{1}{\rho_0^{c_s^2}}[(-g^{rr}-(v^r_0)^2)\partial_r\psi_1-v^r_0v^t_0\partial_t\psi_1]
\end{equation}
Substituting the $v^r,v^t$ and $\rho$ in the continuity Eq. (\ref{continuity_eq_equatorial})  using Eq. (\ref{pert_v}) and (\ref{pert_rho}) respectively and retaining the terms which are of first order in perturbation terms gives
\begin{equation}
\partial_t[\sqrt{-\tilde{g}}H_0(\rho_0 v^t_1+v^t_0\rho_1)]+\partial_r[\sqrt{-\tilde{g}}H_0(\rho_0 v^r_1+v^r_0\rho_1)]=0
\end{equation}
substituting $\rho_1,v^t_1$ and $v^r_1$ in the above equation using Eq. (\ref{rho_1}), (\ref{v_t_up_1_velo}) and (\ref{v_1_velo}) respectively gives
\begin{eqnarray}\label{w_velo_final}
\partial_t[k_1(r)\{-g^{tt}+(v_0^t)^2(1-\frac{1}{c_s^2})\}\partial_t\psi_1]+\partial_t[k_1(r)\{v^r_0v^t_0(1-\frac{1}{c_s^2})\}\partial_r\psi_1]\\ \nonumber
+\partial_r[k_1(r)\{v^r_0v^t_0(1-\frac{1}{c_s^2})\}\partial_t\psi_1]+\partial_r[k_1(r)\{g^{rr}+(v^r_0)^2(1-\frac{1}{c_s^2})\}\partial_r\psi_1]=0 
\end{eqnarray}
where $k_1(r)=-\frac{\sqrt{-\tilde{g}}H_0}{\rho_0^{c_s^2-1}}$. The above equation can be written as $\partial_\mu (f_1^{\mu\nu}\partial_\nu \psi_1)=0$, where $f_1^{\mu\nu}$ is 
\begin{equation}\label{f_velocity}
f_1^{\mu\nu}=k_1(r)\left[
\begin{array}{cc}
-g^{tt}+(v_0^t)^2(1-\frac{1}{c_s^2}) & v^r_0v^t_0(1-\frac{1}{c_s^2})\\
v^r_0v^t_0(1-\frac{1}{c_s^2}) & g^{rr}+(v^r_0)^2(1-\frac{1}{c_s^2})
\end{array}
\right]
\end{equation}
\subsection{Linear perturbation of mass accretion rate}\label{sec:mass}
For stationary flow the $\partial_t$ part of the equation of continuity, i.e., Eq. (\ref{continuity_eq_equatorial}) vanishes and integration over spatial coordinate provides 
\begin{equation}
\sqrt{-\tilde{g}}H_0 \rho_0 v^r_0 ={\rm constant}.
\end{equation}
Multiplying the quantity $\sqrt{-\tilde{g}}H_0\rho_0 v^r_0$ by the azimuthal component of volume element $d\phi$ and integrating the final expression gives the mass accretion rate,
\begin{equation}\label{mass_accretion_rate}
\Psi_0=\Omega\sqrt{-\tilde{g}}H_0\rho_0v^r_0.
\end{equation}
$\Omega$ arises due to the integral over $\phi$ and is just a geometrical factor and therefore can be absorbed in the left hand side to redefine the mass accretion rate without any loss of generality as
\begin{equation}
\Psi_0=\sqrt{-\tilde{g}}H_0\rho_0v^r_0.
\end{equation}
We now define a quantity $\Psi\equiv\sqrt{-\tilde{g}}H \rho(r,t) v^r(r,t)$ which has the stationary value equal to $\Psi_0$. Using the perturbed quantities given by Eq. (\ref{pert_v}) and (\ref{pert_rho})  we have
\begin{equation}\label{pert_mass_accretion_rate}
\Psi(r,t)=\Psi_0+\Psi_1(r,t),
\end{equation}
where
\begin{equation}\label{psi_1}
\Psi_1(r,t)=\sqrt{-\tilde{g}}H_0(\rho_0 v^r_1+v^r_0 \rho_1).
\end{equation}
Using Eq. (\ref{pert_v})-(\ref{del_t_pert_v_up_t}) and (\ref{pert_mass_accretion_rate}) in the continuity Eq. (\ref{continuity_eq_equatorial})  gives
\begin{equation}\label{del_r_psi_1}
a_1\partial_t v^r_1+b_1\partial_t \rho_1=\partial_r \Psi_1,
\end{equation}
where
\begin{equation}\label{a_1_and_b_1}
a_1=-\sqrt{-\tilde{g}}H_0\rho_0\eta_2\quad{\rm and}\quad b_1=-\sqrt{-\tilde{g}}H_0(v^t_0+\rho_0\eta_1).
\end{equation}
Also differentiating Eq. (\ref{psi_1}) with respect to $t$ gives
\begin{equation}\label{del_t_psi_1}
c_1\partial_t v^r_1+d_1\partial_t \rho_1=\partial_t\Psi_1,
\end{equation}
where
\begin{equation}\label{c_1_and_d_1}
c_1=\sqrt{-\tilde{g}}H_0\rho_0\quad{\rm and}\quad d_1=\sqrt{-\tilde{g}}H_0v^r_0.
\end{equation}
With these two equation given by Eq. (\ref{del_r_psi_1}) and (\ref{del_t_psi_1}) we can write $\partial_t v^r_1$ and $\partial_t \rho_1$ solely in terms of derivatives of $\Psi_1$ as
\begin{equation}\label{del_t_rho_1_and_v_1}
\partial_t v^r_1=\frac{1}{\Delta_1}[b_1\partial_t\Psi_1-d_1\partial_r\Psi_1],\quad \partial_t \rho_1=\frac{1}{\Delta_1}[-a_1\partial_t\Psi_1+c_1\partial_r\Psi_1]
\end{equation}
with $\Delta_1=(b_1c_1-a_1d_1)=(\sqrt{-\tilde{g}}H_0)^2\rho_0 \tilde{\Lambda}$ where $\tilde{\Lambda}$ is given by
\begin{equation}\label{Lambda_tilde}
\tilde{\Lambda}=\frac{g_{rr}(v^r_0)^2}{\Lambda v^t_0}-v^t_0+\frac{c_{s0}^2}{\Lambda v^t_0}[\Lambda (v^t_0)^2-1-g_{rr}(v^r_0)^2].
\end{equation}
Now let us go back to the irrotationality condition given by the Eq. (\ref{irrotationality_condition}). Using $\mu=t$ and $\nu=r$ and dividing by $\rho^{c_s^2} v_t$ gives the following equation
\begin{equation}\label{w_mass_1}
\frac{g_{rr}}{v_t}\partial_t v^r+\frac{c_s^2 g_{rr}v^r}{\rho v_t}\partial_t \rho-\partial_r\left(\ln(\rho^{c_s^2}v_t) \right)=0
\end{equation}
Let us substitute the density and velocities using Eq. (\ref{pert_rho}), (\ref{pert_v}) and 
\begin{equation}
\label{pert_v_t_lower}
v_t(r,t)=v_{t0}(r)+v_{t1}(r,t).
\end{equation}
Keeping only the terms that are linear in the perturbed quantities and differentiating with respect to time $t$ gives   
\begin{equation}\label{w_mass_2}
\partial_t[\frac{g_{rr}}{v_{t0}}\partial_t v^r_1 ]+\partial_t[ \frac{g_{rr}c_s^2 v^r_0}{\rho_0 v_{t0}}\partial_t \rho_1]-\partial_r[ \frac{1}{v_{t0}}\partial_t v_{t1}]-\partial_r[ \frac{c_s^2}{\rho_0}\partial_t \rho_1]=0
\end{equation}
We can also write
\begin{equation}\label{del_t_pert_v_lower_t}
\partial_t v_{t1}=\tilde{\eta}_1\partial_t \rho_1+\tilde{\eta}_2 \partial_t v^r_1
\end{equation}
with
\begin{equation}\label{eta_2_tilde_and_eta_2_tilde}
\tilde{\eta}_1=-[\Lambda \eta_1+\frac{g_{\phi t}v_{\phi 0}c_s^2}{g_{\phi\phi}\rho_0} ],\quad\tilde{\eta}_2=-\Lambda \eta_2
\end{equation}
using Eq. (\ref{del_t_pert_v_lower_t}) the Eq. (\ref{w_mass_2}) can be written as
\begin{equation}\label{w_mass_3}
\partial_t[\frac{g_{rr}}{v_{t0}}\partial_t v^r_1 ]+\partial_t[\frac{g_{rr}c_s^2 v^r_0}{\rho_0 v_{t0}}\partial_t \rho_1]-\partial_r[ \frac{\tilde{\eta}_2}{v_{t0}}\partial_t v^r_1]-\partial_r[ (\frac{\tilde{\eta_1}}{v_{t0}}+\frac{c_s^2}{\rho_0})\partial_t \rho_1]=0
\end{equation}
Finally substituting $\partial_t v^r_1$ and $\partial_t \rho_1$ in Eq. (\ref{w_mass_3}) using Eq. (\ref{del_t_rho_1_and_v_1}) we get
\begin{eqnarray}\label{w_mass_final}
\partial_t[ k_2(r)\{-g^{tt}+(v^t_0)^2(1-\frac{1}{c_s^2}) \}\partial_t\Psi_1]+\partial_t[ k_2(r)\{v^r_0v_0^t(1-\frac{1}{c_s^2}) \}\partial_r\Psi_1]\\ \nonumber
+\partial_r [ k_2(r)\{v^r_0v_0^t(1-\frac{1}{c_s^2}) \}\partial_t\Psi_1]+\partial_r [ k_2(r)\{ g^{rr}+(v^r_0)^2(1-\frac{1}{c_s^2})\}\partial_r\Psi_1]=0
\end{eqnarray}
where
\begin{equation}\label{conformal_mass}
k_2(r)=\frac{g_{rr}v^r_0c_s^2}{v^t_0 v_{t0}\tilde{\Lambda}} \quad {\rm and}\quad g^{tt}=\frac{1}{\Lambda}=\frac{1}{g_{tt}+g_{\phi t}^2/g_{\phi\phi}}
\end{equation}
Eq. (\ref{w_mass_final}) can be written as $\partial_\mu (f_2^{\mu\nu}\partial_\nu \Psi_1)=0$ where $f_2^{\mu\nu}$ is given by the symmetric matrix

\begin{eqnarray}\label{f_mass}
f_2^{\mu\nu}=\frac{g_{rr}v^r_0c_s^2}{v^t_0 v_{t0}\tilde{\Lambda}}\left[\begin{array}{cc}
-g^{tt}+(v^t_0)^2(1-\frac{1}{c_s^2}) & v^r_0v_0^t(1-\frac{1}{c_s^2})\\
v^r_0v_0^t(1-\frac{1}{c_s^2}) & g^{rr}+(v^r_0)^2(1-\frac{1}{c_s^2})
\end{array}\right]
\end{eqnarray}

\subsection{Linear perturbation of relativistic Bernoulli's constant}\label{sec:bernoulli}
Energy momentum conservation equation is given by Eq. (\ref{mom_consv_eq}). The energy momentum conservation equation can also be written as $\nabla_\mu T^\mu_{~\nu}=0$. Thus using $\nabla_\mu v_\nu=\partial_\mu v_\nu-\Gamma^\lambda_{\mu\nu}v_\lambda$ we have
\begin{equation}\label{euler_lower}
v^\mu \partial_\mu v_\nu-\Gamma^\lambda_{\mu\nu}v_\lambda v^\mu+\frac{c_s^2}{\rho}(v^\mu v_\nu\partial_\mu \rho+\partial_\nu \rho)=0
\end{equation}
Therefore the temporal component $\nu=t$ of the relativistic Euler equation is given by
\begin{equation}\label{Euler_temporal}
v^t\partial_t v_t+v^r\partial_r v_t-\Gamma^\lambda_{\mu t}v_\lambda v^\mu+\frac{c_s^2}{\rho}(v^t v_t\partial_t\rho+v^rv_t\partial_r\rho+\partial_t\rho)=0.
\end{equation}
It can be shown that $\Gamma^\lambda_{\mu t}v_\lambda v^{\mu}=0$. Thus the Eq. (\ref{Euler_temporal}) can be written as
\begin{equation}\label{euler_mass}
v^t\partial_t v_t+\frac{c_s^2}{\rho}(v^t v_t+1)\partial_t\rho+v^rv_t\partial_r\{\ln(v_t\rho^{c_s^2})\}=0
\end{equation}
For stationary case where all derivatives with respect to $t$ vanish, the solution of the above equation gives the relativistic Bernoulli's constant as $\xi_0=v_{t0}\rho_0^{c_s^2}$ which is a constant of motion for stationary flow.

Let us define a quantity $\xi(r,t)\equiv\rho^{c_s^2}(r,t)v_t(r,t)$ such that it has the stationary value equal to $\xi_0$. Thus
\begin{equation}\label{pert_xi}
\xi(r,t)=\xi_0+\xi_1(r,t)
\end{equation} Now we use the perturbation equations given by Eq. (\ref{pert_rho}), (\ref{pert_v}), (\ref{pert_v_t_lower}) along with Eq. (\ref{pert_xi}) in the Eq. (\ref{euler_mass}) and retain only the terms that are linear in the perturbed quantities. This provides
\begin{equation}\label{del_r_xi_1}
\frac{\xi_0 g_{rr}v^r_0}{v_{t0}}\frac{c_s^2}{\rho_0}\partial_t\rho_1+\frac{\xi_0g_{rr}}{v_{t0}}\partial_t v^r_1=\partial_r\xi_1
\end{equation}
In deriving the above equation we have also used Eq. (\ref{del_t_pert_v_lower_t}), (\ref{eta_2_tilde_and_eta_2_tilde}) and (\ref{eta_1_eta_2_and_Lambda}). $\xi_1(r,t)$ is given by
\begin{equation}\label{xi_1}
\xi_1(r,t)=\frac{c_s^2 \xi_0}{\rho_0}\rho_1(r,t)+\frac{\xi_0}{v_{t0}}v_{t1}
\end{equation}
Differentiating both sides of the above equation with respect to $t$ and using Eq. (\ref{del_t_pert_v_lower_t}) we get
\begin{equation}\label{del_t_xi_1}
\xi_0[\frac{c_s^2}{\rho_0}+\frac{\tilde{\eta}_1}{v_{t0}}]\partial_t \rho_1+\frac{\xi_0 \tilde{\eta}_2}{v_{t0}}\partial_t v^r_1=\partial_t\xi_1
\end{equation}
where $\tilde{\eta}_1$ and $\tilde{\eta}_2$ are given by Eq. (\ref{eta_2_tilde_and_eta_2_tilde}). Eq. (\ref{del_r_xi_1}) and (\ref{del_t_xi_1}) gives $\partial_t v^r_1$ and $\partial_t \rho_1$ solely in terms of derivatives of $\xi_1$
\begin{equation}\label{del_t_pert_v_1_and_rho_1_bernoulli}
\partial_t v^r_1=\frac{1}{\Delta_2}[b_2\partial_t \xi_1-d_2\partial_r\xi_1],\quad \partial_t\rho_1=\frac{1}{\Delta_2}[-a_2\partial_t \xi_1+c_2\partial_r\xi_1]
\end{equation}
where
\begin{equation}\label{a_2_b_2_c_2_d_2}
a_2=\frac{\xi_0g_{rr}}{v_{t0}},\quad b_2=\frac{\xi_0 g_{rr}v^r_0}{v_{t0}}\frac{c_s^2}{\rho_0},\quad c_2=\frac{\xi_0 \tilde{\eta}_2}{v_{t0}},\quad d_2=\xi_0[\frac{c_s^2}{\rho_0}+\frac{\tilde{\eta}_1}{v_{t0}}]
\end{equation}
and
\begin{equation}\label{Delta_2}
\Delta_2=b_2c_2-a_2d_2=\frac{\xi_0^2 g_{rr}c_s^2}{\rho_0 v_{t0}^2v^t_0}
\end{equation}
The continuity equation is given by Eq. (\ref{continuity_eq_equatorial}). Using the perturbation equation given by Eq. (\ref{pert_rho}), (\ref{pert_v}) and retaining only the terms that are linear in perturbed quantities gives
\begin{equation}\label{w_ber_1}
\partial_t[\sqrt{-\tilde{g}}H_{0}(\rho_0v^t_1+v^t_0\rho_1)]+\partial_r[\sqrt{-\tilde{g}}H_0(\rho_0v^r_1+v^r_0\rho_1)]=0
\end{equation}
Differentiating the above equation with respect to $t$ and using Eq. (\ref{del_t_v_up_t_1}) gives
\begin{equation}\label{w_ber_2}
 \fl \partial_t[\sqrt{-\tilde{g}}H_0\{(\rho_0\eta_2)\partial_t v^r_1+(\rho_0\eta_1+v^t_0)\partial_t \rho_1\}]+\partial_r[\sqrt{-\tilde{g}}H_0(\rho_0\partial_t v^r_1+v^r_0\partial_t\rho_1)]=0
\end{equation}
Substituting $\partial_t v^r_1$ and $\partial_t\rho_1$ in the above equation using Eq. (\ref{del_t_pert_v_1_and_rho_1_bernoulli}) provides
\begin{eqnarray}
\label{w_ber_final}
 \partial_t[ k_3(r)\{-g^{tt}+(v^t_0)^2(1-\frac{1}{c_s^2}) \}\partial_t\xi_1
]
+\partial_t[ k_3(r)\{v^r_0v_0^t(1-\frac{1}{c_s^2}) \}\partial_r\xi_1]\\ \nonumber
+\partial_r [ k_3(r)\{v^r_0v_0^t(1-\frac{1}{c_s^2}) \}\partial_t\xi_1]
+\partial_r [ k_3(r)\{g^{rr}+(v^r_0)^2(1-\frac{1}{c_s^2})\}\partial_r\xi_1]=0
\end{eqnarray}
with $k_3(r)=\frac{\sqrt{-\tilde{g}} H_0}{\rho_0^{c_s^2-1}}$. The above equation can be written as $\partial_\mu (f_3^{\mu\nu}\partial_\nu \xi_1)=0$ where $f_3^{\mu\nu}$ is given by the symmetric matrix

\begin{equation}\label{f_bernoulli}
f_3^{\mu\nu}=\frac{\sqrt{-\tilde{g}}H_0}{\rho_0^{c_s^2-1}}\left[\begin{array}{cc}
-g^{tt}+(v^t_0)^2(1-\frac{1}{c_s^2}) & v^r_0v_0^t(1-\frac{1}{c_s^2})\\
v^r_0v_0^t(1-\frac{1}{c_s^2}) & g^{rr}+(v^r_0)^2(1-\frac{1}{c_s^2})
\end{array}\right]
\end{equation}

\section{The acoustic metric}
The linear perturbation of the velocity potential, mass accretion rate  and the relativistic Bernoulli constant, performed in Sec. (\ref{sec:velo}),(\ref{sec:mass}) and (\ref{sec:bernoulli}) respectively, provides equations
\begin{equation}
\partial_\mu(f_i^{\mu\nu}\partial_\nu \tilde{\psi}_i)=0,\quad i=1,2,3;\quad \tilde{\psi}_i=(\psi_1,\Psi_1,\xi_1)
\end{equation}
These equations could be compared to the wave equation of a massless scalar field $\varphi$ propagating in a curved spacetime given by \cite{birrell1984quantum}
\begin{equation}
\partial_\mu (\sqrt{-{g}}g^{\mu\nu}\partial_\nu \varphi)=0
\end{equation}
Thus the acoustic metric $G_i^{\mu\nu}$ can be obtained from the relation\cite{Barcelo}
\begin{equation}
 \sqrt{-G_i}G_i^{\mu\nu}=f_i^{\mu\nu},\quad\quad i=1,2,3
 \end{equation} 
where $G$ is the determinant of the acoustic metric $G_{\mu\nu}$. The $f_i^{\mu\nu}$'s for $i=1,2,3$ are same except the factors $k_i$'s. Thus the acoustic $G_i^{\mu\nu}$ will also be the same up to a conformal factor equal to $\frac{k_i}{\sqrt{-G_i}}$. The acoustic metric obtained by linear perturbing three different quantities differ by only this conformal factor. Thus we can expect the acoustic curvature or the line element of the acoustic spacetime metric to be different for different acoustic spacetime obtained by perturbing different quantities. However, our goal is to investigate the features of the acoustic spacetime which are common to all the acoustic spacetime metric irrespective of the quantity that we have perturbed to obtain the same. The location of the event horizon, causal structure of the spacetime or the surface gravity do not depend on the conformal factor of the spacetime metric. Therefore we are interested in studying these conformally invariant features of the acoustic spacetime. Thus in order to investigate these properties of the acoustic spacetime we could take the acoustic metric to be the same by ignoring the conformal factors. Hence the acoustic metrics $G^{\mu\nu}$ and $G_{\mu\nu}$, ignoring the conformal factor, are given by 
\begin{equation}
G^{\mu\nu}=\left[\begin{array}{cc}
-g^{tt}+(v^t_0)^2(1-\frac{1}{c_s^2}) & v^r_0v_0^t(1-\frac{1}{c_s^2})\\
v^r_0v_0^t(1-\frac{1}{c_s^2}) & g^{rr}+(v^r_0)^2(1-\frac{1}{c_s^2})
\end{array}\right]
\end{equation} 
and
\begin{equation}\label{acoustic_metric}
G_{\mu\nu}=\left[
\begin{array}{cc}
-g^{rr}-(v^r_0)^2(1-\frac{1}{c_s^2}) & v^r_0v_0^t(1-\frac{1}{c_s^2})\\
v^r_0v_0^t(1-\frac{1}{c_s^2}) & g^{tt}-(v^t_0)^2(1-\frac{1}{c_s^2}) 
\end{array} \right]
\end{equation}
\section{Location of the acoustic event horizon}
Eq. (\ref{acoustic_metric}) gives the metric corresponding to the acoustic spacetime. The metric $G_{\mu\nu}$ is stationary. In general relativity, the event horizon of such spacetime is defined as a  hypersurface $r={\rm constant}$ whose normal $n_{\mu}=\partial_\mu r =\delta^r_\mu$ is null with respect to the spacetime metric. Similarly the acoustic event horizon may be defined as a null hypersurface. Thus on the acoustic horizon  we have the condition \cite{Poisson2004relativist,carroll2004spacetime,abraham:causal,Gualtieri2011}. 
\begin{equation}
G^{\mu\nu}n_{\mu}n_\nu = G^{\mu\nu}\delta^r_\mu\delta^r_\nu = G^{rr}=0.
\end{equation}
Therefore on the event horizon we have 
\begin{equation}\label{horizon:1}
c_s^2=\frac{g_{rr}(v^r_0)^2}{1+g_{rr}(v^r_0)^2}.
\end{equation}

In terms of $ u_0 $ (the stationary value of $ u $ given by equation (\ref{v_in_CRF})) the, location of the event horizon is given by the condition 
\begin{equation}
u_0^2|_{\rm h}=c_s^2|_{\rm h}.
\end{equation}
The subscript ``h" implies that the quantity is to be evaluated at the horizon and would imply the same hereafter.
For transonic fluid, the transonic surface is the surface where the advective velocity becomes equal to the sound speed, i.e., where the flow goes to supersonic state from subsonic state. This surface is thus defined by the relation $u_0^2=c_s^2$.
Hence the acoustic event horizon coincides with the transonic surface.

In the following, we show how the acoustic horizon location depends on the Kerr spin parameter $ a $. To do that, first, we set up the equations and parameter space needed to solve the stationary isothermal accretion flow onto a rotating Kerr black hole and identify the position at which $ u_0^2=c_s^2 $.
\subsection{Choice of parameters [$ T,\lambda,a $]}
The characteristic features of the isothermal accretion flow onto a rotating Kerr black hole in the flow model considered here are determined by the set of three parameters $ [T,\lambda,a] $.  Taking logarithmic derivative of the expression for the Bernoulli's constant  and the expression for mass accretion rate (\ref{mass_accretion_rate}) one can find the gradient of the advective velocity to be given by\cite{Tarafdar2016-2}
\begin{equation}\label{advective_gradient}
\frac{du}{dr} = \frac{\frac{1-c_s^2}{2c_s^2}\frac{\mathcal{B}'}{\mathcal{B}}-\frac{1}{2c_s^2}\frac{B'}{B}-\frac{1}{r}}{\frac{1}{u}-\frac{u}{1-u^2}\frac{1-c_s^2}{c_s^2}}\equiv\frac{N}{D}
\end{equation}
where $\mathcal{B}=r^2-2r+a^2$ as defined in Eq. (\ref{A-B-C}) and $ B =  (g_{\phi\phi}+2\lambda g_{\phi t}-\lambda^2g_{tt})$ and prime represents derivative with respect to the radial coordinate $ r $.  The critical points are obtained from the condition $ N=D=0 $. This gives the following condition to be satisfied at the critical points
\begin{equation}\label{critical_points_condition}
u^2_c=c_{sc}^2=\frac{\frac{\mathcal{B}'}{\mathcal{B}}-\frac{B'}{B}}{\frac{2}{r}+\frac{\mathcal{B}'}{\mathcal{B}}}
\end{equation}
The roots of the above equation lying outside the event horizon of the Kerr black hole (given by $ r_+=1+\sqrt{1-a^2} $) are the critical points of the flow.  
Due to the fact that $ N=D=0 $ at the critical points ($ r=r_c $), the advective velocity gradient $ \frac{du_0}{dr}|_c $ at the critical points is obtained by taking the limit as $ r\to r_c $ (and $ u_0^2\to c_{s0}^2 $) by using L'Hospital rule. This gives the velocity gradient to be
\begin{equation}\label{dudr_c}
\frac{du_0}{dr}|_c = \pm\sqrt{\frac{\beta}{\Gamma}}
\end{equation} 
where the plus sign is for wind solution and minus sign is for accretion and  
\begin{equation}\label{Gamma}
\Gamma = \frac{2}{c_{s}^2(1-c_s^2)}
\end{equation}
and
\begin{equation}\label{beta}
\beta = \frac{1-c_s^2}{2c_s^2}\left(\frac{\mathcal{B}'^2}{\mathcal{B}^2}-\frac{\mathcal{B}''}{\mathcal{B}}\right)-\frac{1}{2c_s^2}\left(\frac{B'^2}{B^2}-\frac{B''}{B}\right)-\frac{1}{r_c^2}
\end{equation}
where the double prime stands for the second derivative with respect to the radial coordinate $ r $.
Depending on the values of the parameters $ [T,\lambda,a] $, the equation (\ref{critical_points_condition}), and hence the accretion flow, would exhibit either one or three critical points. These critical points are typically shown in the phase portrait of the flow as inner critical point ($ r_{\rm in} $), middle critical point ($ r_{\rm mid} $) and outer critical point ($ r_{\rm out} $), where $ r_{\rm in}<r_{\rm mid}<r_{\rm out} $ (see Fig. \ref{Fig:phase_portrait} for example). By performing a critical point analysis one can show that for a single critical point it must be a saddle type, whereas for three critical points $ r_{\rm in} $ and $ r_{\rm out} $ are saddle type and $ r_{\rm mid} $ is center type. For accretion flow, phase portrait containing only one critical point, the accretion flow must be mono-transonic, for which the accretion flow starts from large radial distance subsonically and becomes supersonic at some radius which is called the transonic point, and enters the black hole supersonically. However, we are more interested in accretion flows which contain multiple critical points. For such flow, it is possible for the flow to start subsonically at large distance and become supersonic at some radius (outer critical point $r_{{\rm out}}$) and then encounter a shock to become subsonic which again becomes supersonic at some smaller radial distance (inner critical point $r_{{\rm in}}$). Thus such flow is multi-transonic. We would like to use such parameter set which allows multiple critical points. However multi-critical flows are not necessarily multi-transonic. For a multi-critical flow to become multi-transonic it must encounter a shock \cite{TKD2012-Czerny}. But all parameters which allow multiple critical points do not allow shock formation. Thus the parameters space which allows shock formation is smaller than the parameter space for multiple critical points.

To allow shock formation the integral solution must satisfy the shock conditions i.e the  general relativistic
Rankine Hugoniot conditions. These conditions are given by 
\begin{equation}\label{R-H_condition}
	[[\rho v^\mu]]=0,\quad{\rm and} [[T^{\mu\nu}]]=0
\end{equation}
where $[[V]]=V_+-V_-$, $V_-$ and $V_+$ symbolically denoting the values of some
flow variable $V$ before and after the shock respectively. Location of shock formation is obtained by equating the \textit{shock-invariant quantity} $S_{h}$ on the integral solution passing through the outer critical point to that on the integral solution passing through the inner critical point exclusively at the shock location. The \textit{shock-invariant quantity} $S_{h}$ for isothermal conical flow in Kerr spacetime is obtained to be \cite{Yuan1996,Tarafdar2016-2}
\begin{equation}\label{S_h}
	S_h=\left(\frac{u_0}{\sqrt{1-u_0^2}} \right)^{2c_s^2-1}(u_0^2\mathcal{B}+r^2c_s^2(1-u_0^2))
\end{equation}
 For a particular solution described by $[T,\lambda,a]$, to find the shock location numerically we evaluate $S_h$ on the supersonic part of the solution passing through the outer critical point and on the subsonic part of the solution passing through the inner critical point at each $r$. The value of $r$ where these two values match is taken to be the location at which the stationary shock forms. 

Before we construct the parameter space which allows multi-transonic flow, we make the following considerations. We take the specific angular momentum $ \lambda $ to be positive and the black hole spin $ a>0 $ for prograde motion and $ a<0 $ for retrograde motion. We assumed the flow to be inviscid. This restricts the value of $ \lambda $ to be low in order that accretion happen. As mentioned earlier, for a large value of $ \lambda $, the fluid may settle into stable Keplerian orbits around the black hole and in order that the accretion happens viscosity must play a role to transport the angular momentum outward helping the matter to fall radially inwards. While constructing the $ [a,\lambda] $ parameter space we concentrate only on those values of $ \lambda $ for which accretion flow remains sub-Keplerian .  The black hole spin have the range $ -1\leq a \leq 1 $.
Let us now consider the bulk ion temperature $ T $. The isothermal sound speed may be expressed in terms of the bulk ion temperature $T$ through the Clayperon-Mendeleev equation \cite{bazarov1964thermodynamics,gibbs2014elementary}
\begin{equation}\label{Clayperon}
	c_s = \sqrt{\frac{k_B}{\mu m_H}T}
\end{equation}
where $k_B$ is the Boltzmann constant, $m_H$ is the mass of the hydrogen atom and $\mu$ is the mean molecular weight. From the above expression we see that for a smaller value of $ T $, the sound speed will also be small.  The Mach number $\mathcal{M}$ is defined  as the ratio of the advective velocity $u_0$ and the sound speed $c_s$, i.e., $\mathcal{M}=\frac{u_0}{c_s}$. At large distance $ u_0 $ is small. As the fluid moves inwards $ u_0 $ increases and at some radius (transonic point, $ r_{\rm out} $) it becomes equal to the speed of sound, i.e., $ \mathcal{M} $ becomes $ 1 $. For isothermal accretion $ c_s $ is constant. Thus for a smaller value of $ c_s $, $ r_{\rm out} $ will be larger and for a larger value of $ c_s $, $ r_{\rm out} $ will be smaller. 

\begin{figure}[H]
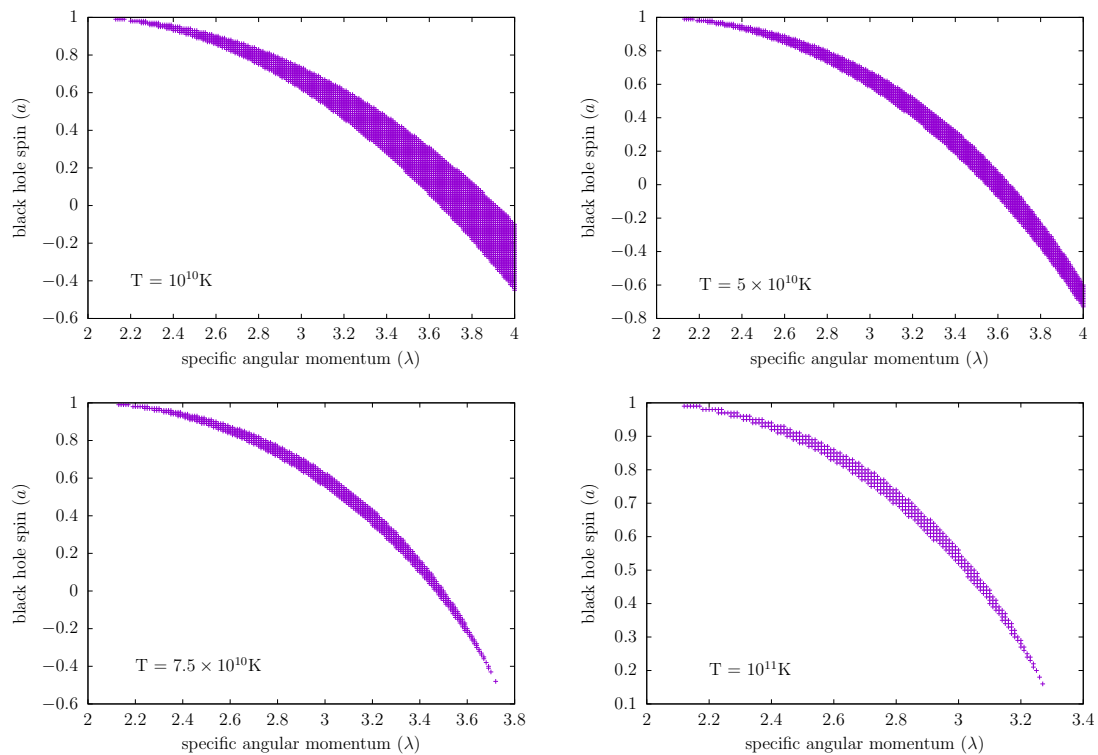

	\centering
	\begin{tabular}{cc}
		\resizebox{0.45\textwidth}{!}{\input{parameterspaceT1}} &
		\resizebox{0.45\textwidth}{!}{\input{parameter_space_T5}}\\
		\resizebox{0.45\textwidth}{!}{\input{parameter_space_T75}}&
		\resizebox{0.45\textwidth}{!}{\input{parameter_space_T11}}
	\end{tabular}
	\caption{Region of parameter space [$ \lambda,a $] allowing multiple critical points for different values of $ T $. For prograde motion higher spin corresponds to lower $ \lambda $ whereas for retrograde motion higher spin corresponds to higher $ \lambda $ for a particular $ T $. For fixed fixed values of $ [T,\lambda] $ only a finite range of $ a $ allow multi-critical accretion.}\label{Fig:parameter_space_T}
\end{figure}

 In Fig. \ref{Fig:parameter_space_T} we show the region of the parameter space $ [a,\lambda] $ which allows multi-critical flow, for four fixed values of temperatures. 
 It can be noticed that for multi-critical accretion at a fixed value of $ T $, for prograde motion higher spin corresponds to lower $ \lambda $ whereas for retrograde motion higher spin corresponds to higher $ \lambda $. Here as well as in the following whenever we talk about the dependence of any quantity on the black hole spin $ a $ and explicitly mention whether the motion is prograde or retrograde, we would imply the dependence on the magnitude of the spin $ a $.

\begin{figure}[H]
	\centering
	\begin{tabular}{cc}
		\resizebox{0.45\textwidth}{!}{\input{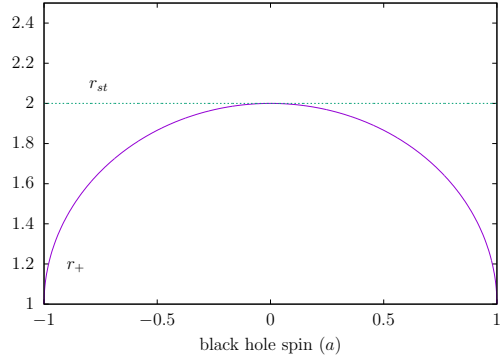}} &
	\end{tabular}
	\caption{ $ r_+ $ and $ r_{\rm st} $ in the equatorial plane is plotted along the vertical axis against the black hole spin $ a $ along the horizontal axis, where $ r_{+} $ is the radius of the event horizon and $ r_{\rm st} $ is the radius of the static limit surface. In units $ G=c=M=1 $. }\label{Fig:horizon}
\end{figure}

Before we discuss the dependence of the acoustic horizon on the black hole spin, it would be nice to see how the event horizon of the Kerr black hole and the static limit surface varies with the black hole spin on the equatorial plane. Fig. \ref{Fig:horizon} shows the variation of location of the event horizon ($ r_+ $) and the static limit surface ($ r_{\rm st} $) of the Kerr black hole as a function of the black hole spin $ a $.

\subsection{Dependence of the acoustic horizon on the parameters $ [T,\lambda,a] $}
For fixed values of $ [T,\lambda] $ only a finite range of $ a $ is allowed for multi-critical accretion flow. Therefore we take a set of values of $ [T,\lambda] $ and see the dependence of the acoustic horizon  on the allowed range of $ a $. As the inner acoustic horizon ($ r_{\rm in} $) may be very near to the black hole horizon $ (r_{+}) $ it would also be interesting to see the location of the $ r_{\rm in} $ compared to the radius of the circular orbits of particles, i.e., the photon orbit ($ r_{\rm photon} $), the innermost unstable bound orbit ($ r_{\rm bound} $) and the innermost stable circular orbit ($ r_{\rm ISCO} $), around Kerr black hole  \cite{Bardeen1972,chandrasekhar1983mathematical,Novikov2012black}. The radius of innermost circular orbit, closest to the black hole, (along which the motion is at the speed of light) is given by
\begin{equation}\label{photn_orbit}
r_{\rm photon} = 2 \left(1+\cos \left(\frac{2}{3} \cos ^{-1}(-a)\right)\right)
\end{equation}
and the last bound but unstable circular orbit on which energy of the orbit is equal to the rest mass of the particle is given by
\begin{equation}\label{bind_orbit}
r_{\rm bound} = 2 - a + 2 \sqrt{1-a}
\end{equation}
if  a particle in the equatorial plane comes from infinity, with $ v_{\infty} \ll c $, where $ c $ is the speed of light, and passes within $ r_{\rm bound} $, then it will be captured by the black hole. Lastly the radius of the inner most circular orbit which is stable is given by
\begin{equation}\label{isco}
r_{\rm ISCO} = 3 + Z_2\mp \sqrt{(3-Z_1)(3+Z_1+2Z_2)}
\end{equation}
where
\begin{equation}
Z_1 = 1+(1-a^2)^{\frac{1}{3}}[(1+a)^{\frac{1}{3}}+(1-a)^{\frac{1}{3}}]
\end{equation}
and 
\begin{equation}
Z_2 = \sqrt{3a^2+Z_1^2}
\end{equation}
the upper sign and lower sign in Eq. (\ref{isco}) are for prograde and retrograde motion of the particle.

\begin{figure}[H]
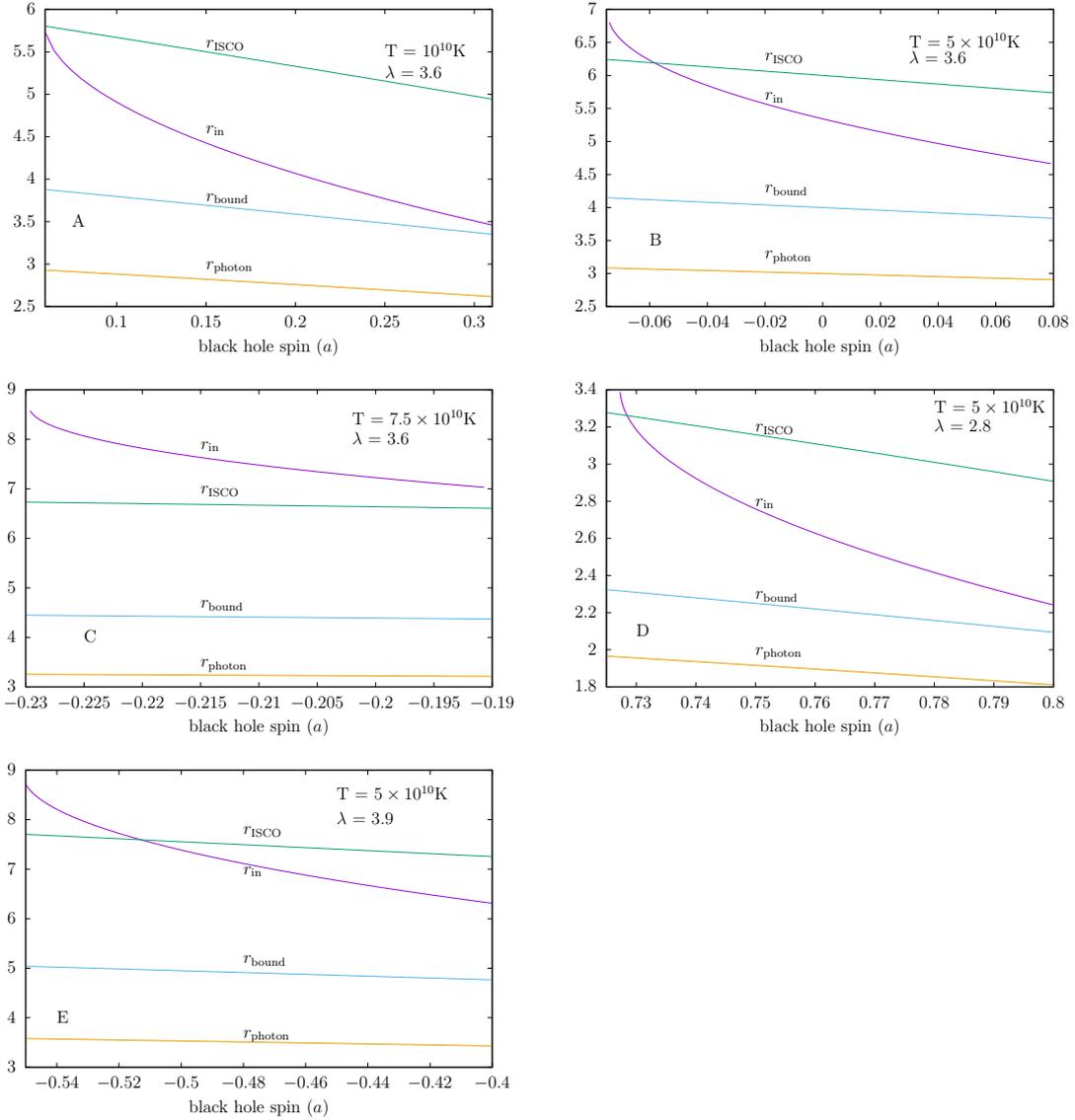

	\centering
	\begin{tabular}{cc}
		\resizebox{0.45\textwidth}{!}{\input{acoustic_h_l36_T10_new}} &
		\resizebox{0.45\textwidth}{!}{\input{acoustic_h_l36_T105_new}}\\
		\resizebox{0.45\textwidth}{!}{\input{acoustic_h_l36_T1075_new}}&
		\resizebox{0.45\textwidth}{!}{\input{acoustic_h_l28_T1050_new}}\\
		\resizebox{0.45\textwidth}{!}{\input{acoustic_h_l39_T1050_new}} &
	\end{tabular}
	\caption{Inner acoustic horizon $ r_{\rm in}  $ (along the vertical axis) vs black hole spin $ a $ (along the horizontal axis) plot for different set of $ [T,\lambda] $ values. $ T $ and $ \lambda $ are the temperature and the specific angular momentum respectively. \textbf{A, B} and \textbf{C} corresponds to a fixed $ \lambda=3.6 $ and $ T = 10^{10}K,5\times10^{10}K $ and $ 7.5\times 10^{10}K $ respectively. \textbf{D, B} and \textbf{E} corresponds to a fixed $ T = 5\times10^{10}K $ and $ \lambda=2.8,3.6 $ and $ 3.9 $ respectively. For all the set of parameters $ [T,\lambda] $, $ r_{\rm in} $ decreases with $ a $ for prograde motion whereas it increases with  $ a $ for retrograde motion. $ r_{\rm photon} $, $ r_{\rm bound} $ and $ r_{\rm ISCO} $ represent the radius of the circular photon orbit, innermost bound but unstable circular  orbit and the innermost stable circular  orbit (ISCO), respectively. At least for these values of parameters $ [T,\lambda] $, inner acoustic horizon $ r_{\rm in} $ always remains outside the innermost unstable bound circular orbit $ r_{\rm bound} $. However, in order to check whether this is true for all the points of the parameter space $ [T,\lambda] $, the analysis should be done for all values of $ [T,\lambda] $. The variation of $ r_{\rm photon} $, $ r_{\rm bound} $ and $ r_{\rm ISCO} $ (given by Eq. (\ref{photn_orbit}), (\ref{bind_orbit}) and (\ref{isco}) respectively) with $ a $ looks almost linear due to the very small range of $ a $ over which they are plotted.}  \label{Fig:acoustic_horizon_inner}
\end{figure}

 Fig. \ref{Fig:acoustic_horizon_inner} shows the dependence of $ r_{\rm in} $ on the black hole spin $ a $ for different set of values of temperature and  specific angular momentum as well as the variation of the radius of the circular orbits with $ a $ for that range .  Plots \textbf{A}, \textbf{B} and \textbf{C} shows the variation of $ r_{\rm in} $ with $ a $ for temperatures $ 10^{10}K $, $ 5\times10^{10} K$ and $ 7.5\times 10^{10} K $ respectively, for a fixed specific angular momentum $ \lambda = 3.6 $. Whereas plots \textbf{D}, \textbf{B} and \textbf{E} shows the same for specific angular momentum $ \lambda = 2.8 $, $ \lambda = 3.6 $ and $ \lambda = 3.9 $ respectively for a fixed temperature $ 5\times 10^{10}K$. 
 For a fixed $ [T,\lambda] $, the location of the inner critical points, i.e., $ r_{\rm in} $ decreases with $ a $ for prograde motion whereas it increases with  $ a $ for retrograde motion. Also by plotting the radius of different circular orbits for the corresponding range of $ a $ it is noticed that at least for these values of parameters $ [T,\lambda] $, inner acoustic horizon $ r_{\rm in} $ always remains outside the innermost unstable bound circular orbit $ r_{\rm bound} $. However in order to check whether this is true for all the points of the parameter space $ [T,\lambda] $ the analysis should be done for all values of $ [T,\lambda] $ which is beyond the scope of the present work.

\begin{figure}[H]
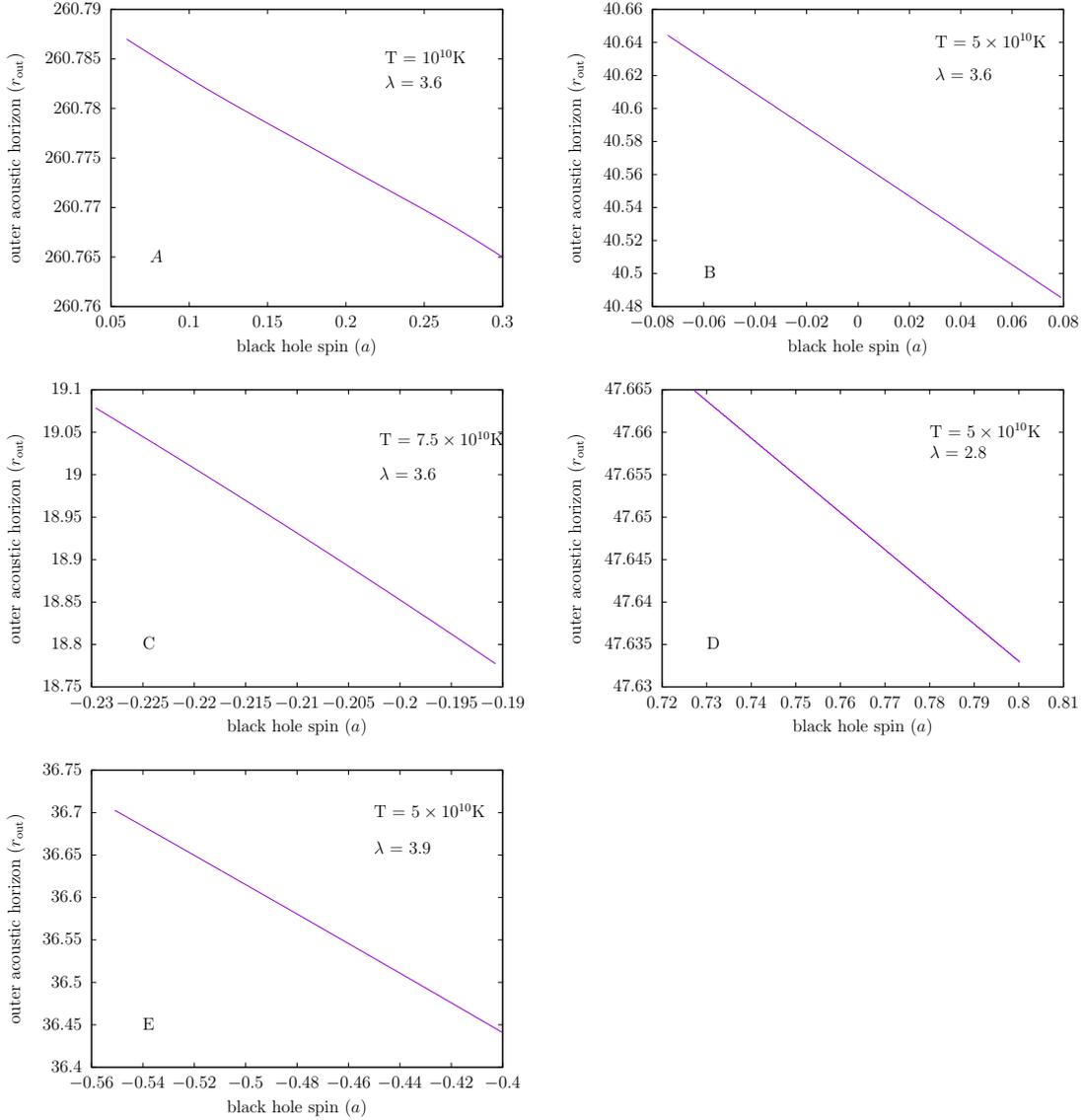

	\centering
	\begin{tabular}{cc}
		\resizebox{0.45\textwidth}{!}{\input{acoustic_h_l36_T10_outer}} &
		\resizebox{0.45\textwidth}{!}{\input{acoustic_h_l36_T105_outer}}\\
	  	\resizebox{0.45\textwidth}{!}{\input{acoustic_h_l36_T1075_outer}}&
		\resizebox{0.45\textwidth}{!}{\input{acoustic_h_l28_T1050_outer}}\\
		\resizebox{0.45\textwidth}{!}{\input{acoustic_h_l39_T1050_outer}}&
	\end{tabular}
	\caption{Outer acoustic horizon $ r_{\rm out} $ vs black hole spin $ a $ plot for different set of $ [T,\lambda] $ values. $ T $ and $ \lambda $ are the temperature and the specific angular momentum respectively. \textbf{A, B} and \textbf{C} corresponds to a fixed $ \lambda=3.6 $ and $ T = 10^{10}K,5\times10^{10}K $ and $ 7.5\times 10^{10}K $ respectively. \textbf{D, B} and \textbf{E} corresponds to a fixed $ T = 5\times10^{10}K $ and $ \lambda=2.8,3.6 $ and $ 3.9 $ respectively. For all the set of parameters $ [T,\lambda] $, $ r_{\rm out} $ decreases as the black hole spin $ a $ increases for prograde motion and increases with increasing $ a $ for retrograde motion.}  \label{Fig:acoustic_horizon_outer}
\end{figure}

Fig. \ref{Fig:acoustic_horizon_outer} shows the dependence of the outer acoustic horizon $ r_{\rm out} $ on the black hole spin $ a $ for different set of values of temperature and  specific angular momentum.  Plots \textbf{A}, \textbf{B} and \textbf{C} shows the variation of $ r_{\rm out} $ with $ a $ for temperatures $ 10^{10}K $, $ 5\times10^{10} K$ and $ 7.5\times 10^{10} K $ respectively, for a fixed specific angular momentum $ \lambda = 3.6 $. Whereas plots \textbf{D}, \textbf{B} and \textbf{E} shows the same for specific angular momentum $ \lambda = 2.8 $, $ \lambda = 3.6 $ and $ \lambda = 3.9 $ respectively for a fixed temperature $ 5\times 10^{10}K$. Here also for a fixed $ [T,\lambda] $, the location of the outer critical points, i.e., $ r_{\rm out} $ decreases as the black hole spin $ a $ increases for prograde motion and increases with increasing $ a $ for retrograde motion. 

From plots \textbf{A}, \textbf{B} and \textbf{C} in Fig. \ref{Fig:acoustic_horizon_outer}, i.e, for same $ \lambda $ (here for $ \lambda=3.6 $), it is seen that for smaller temperature $r_{\rm out}  $ is large compared to that for higher temperature as argued earlier. Also the change in $ r_{\rm out} $ w.r.t $ a $ is quite small as compared to the variation of $ r_{\rm in} $ with $ a $.
This is due to the fact that for large value of $ r_{\rm out} $, the spacetime becomes asymptotically flat, and in such Newtonian gravity limit, the influence of the spin of black hole on the matter flow is less important and that is why change of $ r_{\rm out} $  with respect to $ a $ is small compared to that of $ r_{\rm in} $.

\section{Causal structure of the acoustic spacetime}\label{Sec:Causal_structure}
Acoustic null geodesic corresponding to the radially traveling ($d\phi=0,d\theta=0$) acoustic phonons is given by 
$ds^2=0$. Thus 
\begin{equation}\label{drdt_pm}
(\frac{dr}{dt})_\pm\equiv b_\pm=\frac{-G_{rt}\pm \sqrt{G_{rt}^2-G_{rr}G_{tt}}}{G_{rr}}
\end{equation}
So $t(r)$ can be obtained as 
\begin{equation}\label{t_pm}
t(r)_\pm=t_0+\int_{r_0}^r \frac{dr}{b_\pm}
\end{equation}
We can introduce two new sets of coordinates as following
\begin{equation}\label{dudw}
dz = dt-\frac{1}{b_+}dr,\quad{\rm and},\quad dw = dt-\frac{1}{b_-}dr
\end{equation}
In terms of these new coordinates the acoustic line element can be written as
\begin{equation}
ds^2|_{\phi=\theta={\rm const}}=\mathcal{D}dzdw
\end{equation}
Where $\mathcal{D}$  is found to be equal to $G_{tt}$.
The acoustic metric elements $G_{tt},G_{rt}=G_{tr},G_{rr}$ given by Eq. (\ref{acoustic_metric}) are expressed in terms of the background metric elements and the velocity variables $u_0(r)$ and $\lambda$ using Eq. (\ref{v_in_CRF}) and (\ref{vt_0}). Thus $b_\pm(r)$ is function of the stationary solution $u_0(r)$. Therefore we have to first obtain $u_0(r)$ by solving the relativistic Euler equation for steady state. This is done by numerically integrating equation (\ref{advective_gradient}) which provides the gradient of the advective velocity, i.e, $ \frac{du_0}{dr} $ . We use $4^{\rm {th}}$ order Runge-Kutta method to integrate this equation with the initial condition given by equation (\ref{critical_points_condition}) . Thus for a particular set of $[T,\lambda,a]$ we get $u_0(r)$ numerically and hence we get $b_\pm(r)$. The integration in Eq. (\ref{t_pm}) is then performed by applying Euler method. Finally we plot $t(r)_\pm$ as function of $r$ to see the causal structure of the acoustic spacetime. 

\begin{figure}[H]
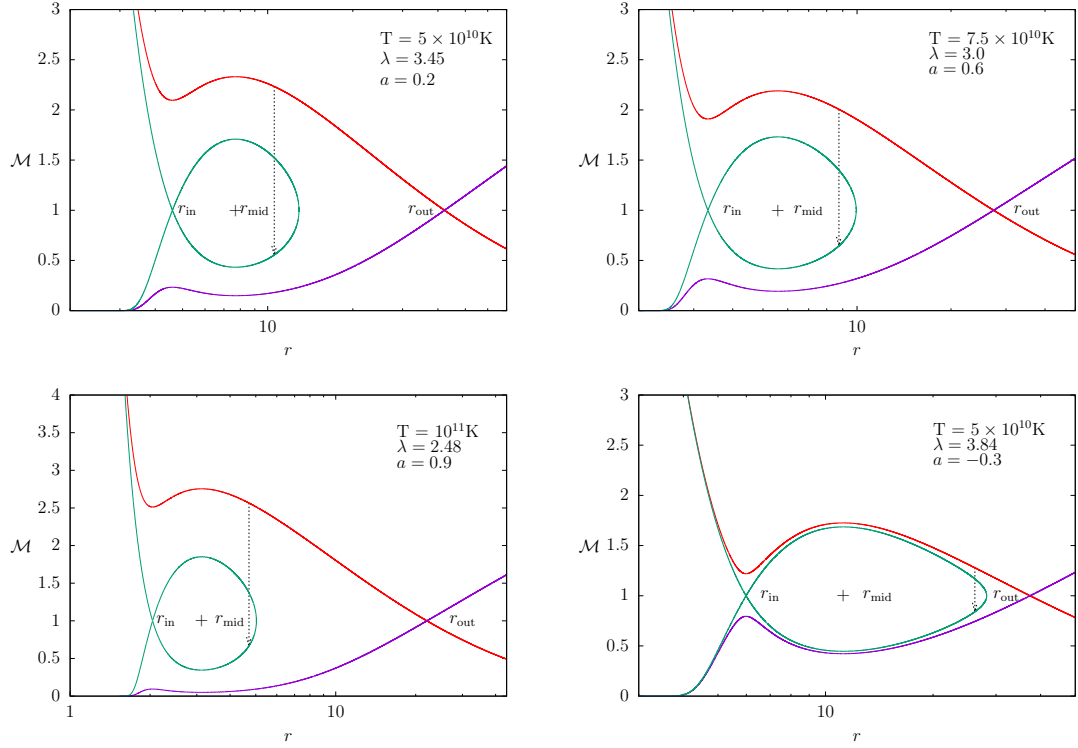

\centering
\begin{tabular}{cc}
 \resizebox{0.45\textwidth}{!}{\input{phase_portrait_a02}} &  \resizebox{0.45\textwidth}{!}{\input{phase_portrait_a06}}\\
 \resizebox{0.45\textwidth}{!}{\input{phase_portrait_a09}} & \resizebox{0.45\textwidth}{!}{\input{phase_portrait_a-03}}
\end{tabular}
\caption{Mach number $\mathcal{M}=\frac{u_0}{c_s}$ vs radial distance $ r $ plots for  different values of $ [T,\lambda,a] $. $ T,\lambda $ and $ a $ are the temperature, the specific angular momentum and the black hole spin respectively. The parameter values for the plots are (row wise) $ [T,\lambda,a]=[5\times10^{10}K,3.45,0.2],[7.5\times10^{10}K,3.0,0.6],[10^{11}K,2.48,0.9] $ and $ [5\times10^{10}K,3.84,-0.3] $ respectively.  The vertical dashed line represents the location of shock formation and the transition from supersonic to subsonic state.}\label{Fig:phase_portrait}
\end{figure}

\begin{figure}[H]
\centering
\begin{tabular}{cc}
	\includegraphics[scale=0.5]{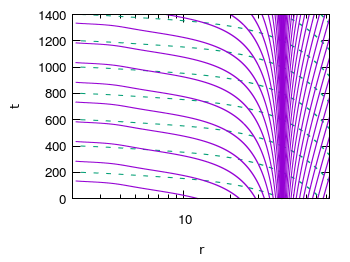} & 
	\includegraphics[scale=0.5]{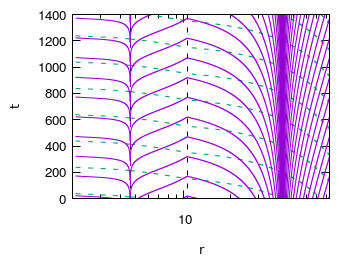} \\
	\includegraphics[scale=0.5]{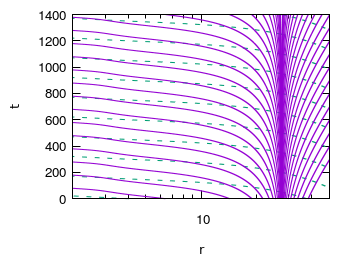} &
	\includegraphics[scale=0.5]{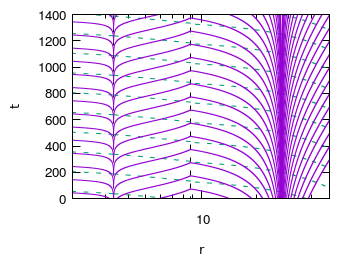} \\ 
	\includegraphics[scale=0.5]{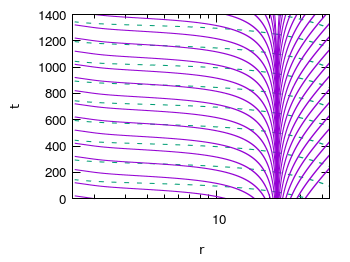} &
	\includegraphics[scale=0.5]{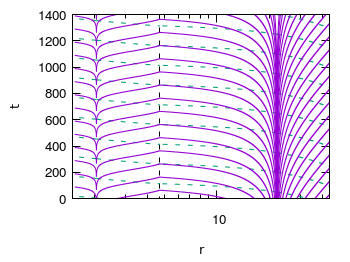} \\
	\includegraphics[scale=0.5]{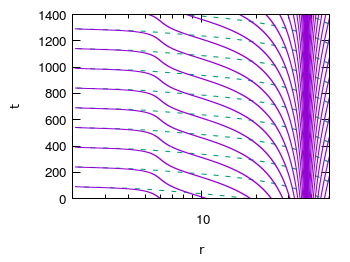} &
	\includegraphics[scale=0.5]{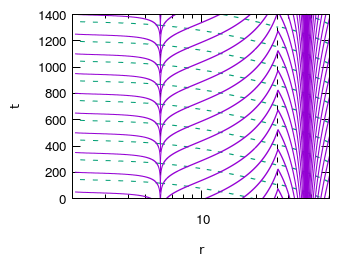} \\
\end{tabular}
\caption{Left: The acoustic causal structures without shock. Right: The acoustic causal structure with shock. The dashed (green) line denotes $t(r)_-$ vs $r$ i.e $w={\rm const}$ and the solid (violet) line denotes $t(r)_+$ vs $r$ i.e $z={\rm const}$. The vertical dashed (black) lines indicates the location of inner acoustic horizon (at smaller radii) and that of the shock formation (larger radii). The rows from top to bottom corresponds to $[T,\lambda,a]=[5\times10^{10}K,3.45,0.2],[7.5\times10^{10}K,3.0,0.6],[10^{11}K,2.48,0.9] $ and $ [5\times10^{10}K,3.84,-0.3] $ respectively.}\label{Fig:causal-structure-shock}
\end{figure}

We take few representative values of the parameters as $[T,\lambda,a]$ and study the causal structure of the corresponding accretion flow. The phase portraits of the stationary solution for four different sets of $[T,\lambda,a]$
are shown in Fig.\ref{Fig:phase_portrait}. The transonic accretion flow passes through the outer critical point $r_{\rm out}$ and becomes supersonic from subsonic state. We obtain the causal structure corresponding to this transonic solution. The $ [T,\lambda] $ are chosen such that the accretion flow allows multiple critical points (See Fig. \ref{Fig:parameter_space_T}) as well as shock formation for the black hole spin $ a=0.2,0.6,0.9 $ and $ -0.3 $. The parameter values for the plots are (row wise) $ [T,\lambda,a]=[5\times10^{10}K,3.45,0.2],[7.5\times10^{10}K,3.0,0.6],[10^{11}K,2.48,0.9] $ and $ [5\times10^{10}K,3.84,-0.3] $, respectively. The vertical dashed line represents the location of shock formation and transition from supersonic state to subsonic state. In Fig. \ref{Fig:causal-structure-shock} we plot the causal structures for the accretion flows ( plotted in Fig. \ref{Fig:phase_portrait}) without shock (left) and with shock (right) . The rows from top to bottom corresponds to $[T,\lambda,a]=[5\times10^{10}K,3.45,0.2],[7.5\times10^{10}K,3.0,0.6],[10^{11}K,2.48,0.9] $ and $ [5\times10^{10}K,3.84,-0.3] $ respectively.

We have also presented here the value of specific angular momentum of test particle around Kerr black hole in different circular orbits for those particular value of spin $ a $ for which we construct the causal structure. In order that accretion (in case of inviscid flow) happen the specific angular momentum of the fluid should not be  higher than that of the test particle in innermost stable circular orbit (ISCO). The specific angular momentum of a circular orbit is given by \cite{Bardeen1972}
\begin{equation}\label{ang_cir_orbit}
\lambda(r,a) = \frac{r^2-2a\sqrt{r}+a^2}{r^{\frac{3}{2}}-2\sqrt{r}+a}
\end{equation}

 In table \ref{table}, we list the radii of the three circular orbits (given by Eq. (\ref{photn_orbit}-\ref{isco})) and the corresponding values of the specific angular momenta.
 
\begin{figure}[H]
	\centering
	\begin{tabular}{c}
		\resizebox{0.45\textwidth}{!}{\input{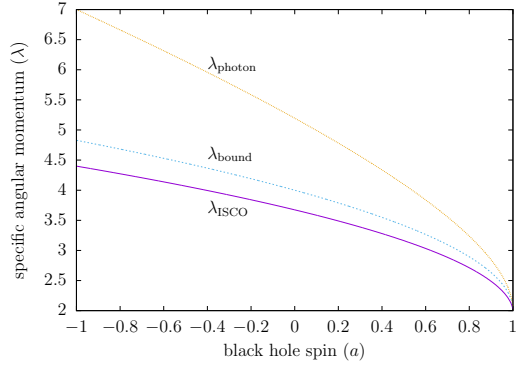}}
	\end{tabular}
	\caption{Specific angular momentum of different circular orbits vs. black hole spin.  $ \lambda_{\rm ISCO} $, $ \lambda_{\rm bound} $ and $ \lambda_{\rm photon} $ are the specific angular momentum of innermost stable circular orbit, innermost unstable bound orbit and circular photon orbit, respectively.}\label{Fig:lambda}
\end{figure}

	\begin{table}[H]
		\centering
		\begin{tabular}{|c|cccc|}\hline
			& $ a=0.2 $ & $ a=0.6 $ & $ a=0.9 $ & $a=-0.3 $ \\ \hline
			$ r_{\rm ISCO} $ & 5.32944 & 3.82907& 2.32088 &  6.94927 \\ 
			$ \lambda_{\rm ISCO} $ & 3.48958 & 3.0326 & 2.48717 & 3.91967 \\ \hline
			$ r_{\rm bound} $ & 3.58885 & 2.66491 & 1.73246 & 4.58035 \\ 	
			$ \lambda_{\rm bound} $ & 3.78885 & 3.26491 &  2.63245 & 4.28035\\ 	\hline
			$ r_{\rm photon} $ & 2.75919 & 2.18891 & 1.55785 & 3.32885  \\ 
			$ \lambda_{\rm photon} $  & 4.78325 & 3.83849 & 2.84442 & 5.77354\\ \hline
		\end{tabular}
		\caption{ Radius and corresponding specific angular momentum for different circular orbits around Kerr black hole for selected values of black hole spin $ a $ for which we construct the causal structures. The specific angular momentum of the fluid $ \lambda $ is taken to be smaller than $ \lambda_{\rm ISCO} $ such that inviscid accretion can happen.}\label{table}
	\end{table}
In Fig. \ref{Fig:lambda} we plot the specific angular momentum of different circular orbits vs. black hole spin.  $ \lambda_{\rm ISCO} $, $ \lambda_{\rm bound} $ and $ \lambda_{\rm photon} $ are the specific angular momentum of innermost stable circular orbit, innermost unstable bound orbit and circular photon orbit, respectively. It should be mentioned that for fluid the specific angular momentum of the  innermost stable circular orbit is not expected to be the same as that of a test particle. Because a fluid element experiences a pressure gradient which is absence in case of a test particle.

As is obvious from the Fig. \ref{Fig:causal-structure-shock}, the shock formation in multi-transonic black hole accretion flow can thus be considered as the presence of an acoustic white hole in the corresponding sonic geometry. Where as the inner and outer transonic surfaces act as acoustic black hole horizons.

The aforementioned procedure to construct the relevant causal structures are based on the assumptions that 
the stationary integral flow solutions are obtained for the steady state (through the integration of the 
time independent Euler and the continuity equations). Such assumptions, however, are to be justified by 
showing that the steady states are stable for this case. In subsequent sections, we thus perform the 
linear stability analysis of the accretion flow to ensure that the steady states are stable states. 

\section{Acoustic surface gravity}
\begin{figure}[H]
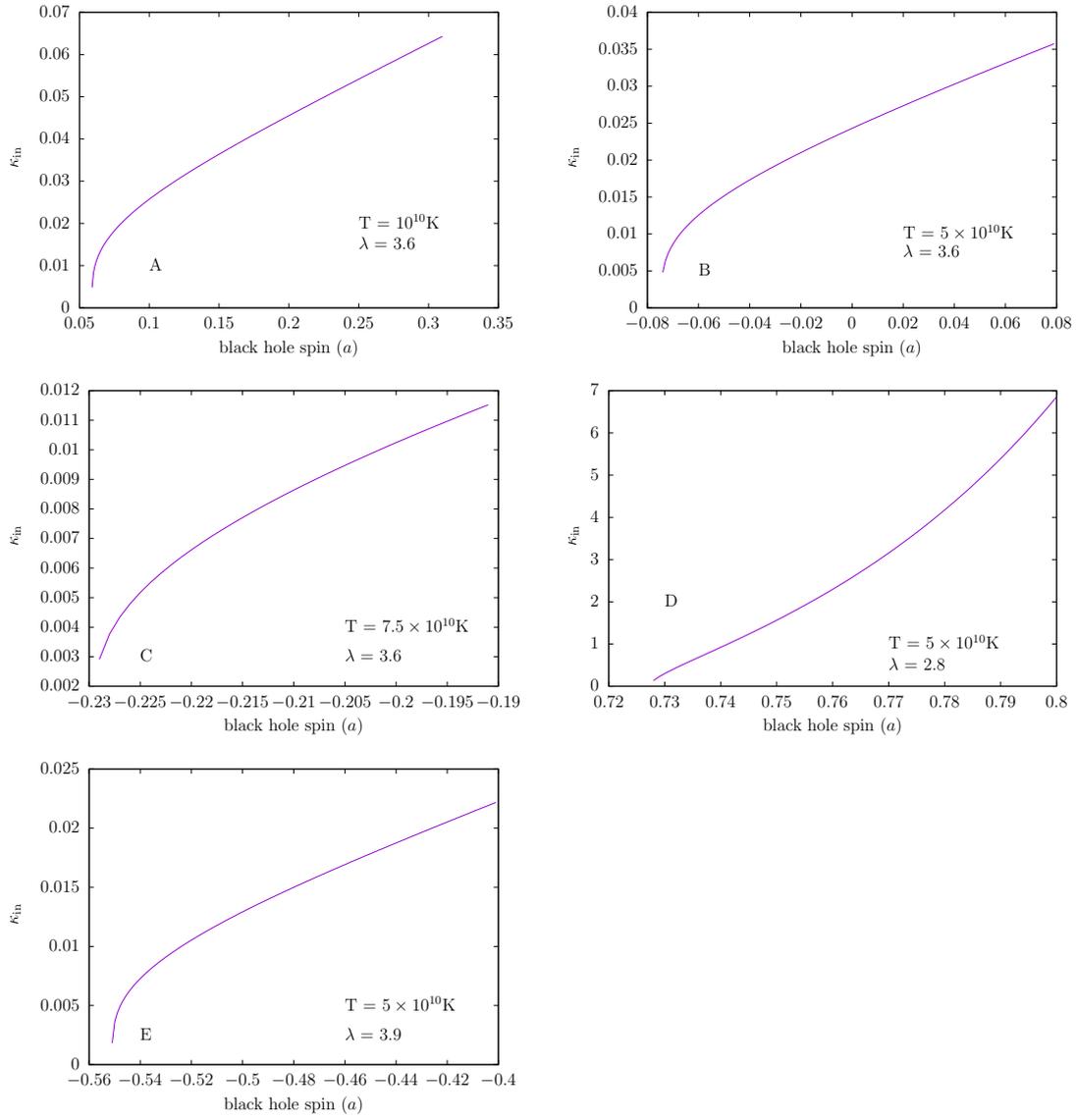

	\centering
	\begin{tabular}{cc}
		\resizebox{0.45\textwidth}{!}{\input{surface_g_l36_T10_in}} & 
		\resizebox{0.45\textwidth}{!}{\input{surface_g_l36_T105_in}}\\ 
		\resizebox{0.45\textwidth}{!}{\input{surface_g_l36_T1075_in}} & 
		\resizebox{0.45\textwidth}{!}{\input{surface_g_l28_T105_in}}\\
		\resizebox{0.45\textwidth}{!}{\input{surface_g_l39_T105_in}}
	\end{tabular}
	\caption{Acoustic surface gravity at the inner horizon ($ \kappa_{\rm in} $) vs black hole spin ($ a $) plot. \textbf{A, B} and \textbf{C} corresponds to a fixed $ \lambda=3.6 $ and $ T = 10^{10}K,5\times10^{10}K $ and $ 7.5\times 10^{10}K $ respectively. \textbf{D, B} and \textbf{E} corresponds to a fixed $ T = 5\times10^{10}K $ and $ \lambda=2.8,3.6 $ and $ 3.9 $ respectively. For all the set of parameters $ [T,\lambda] $, $ \kappa_{\rm in} $ increases with increasing $ a $ for prograde motion and decreases with increasing $ a $ for retrograde motion.}\label{Fig:surface_g_inner}
\end{figure}

\begin{figure}[H]
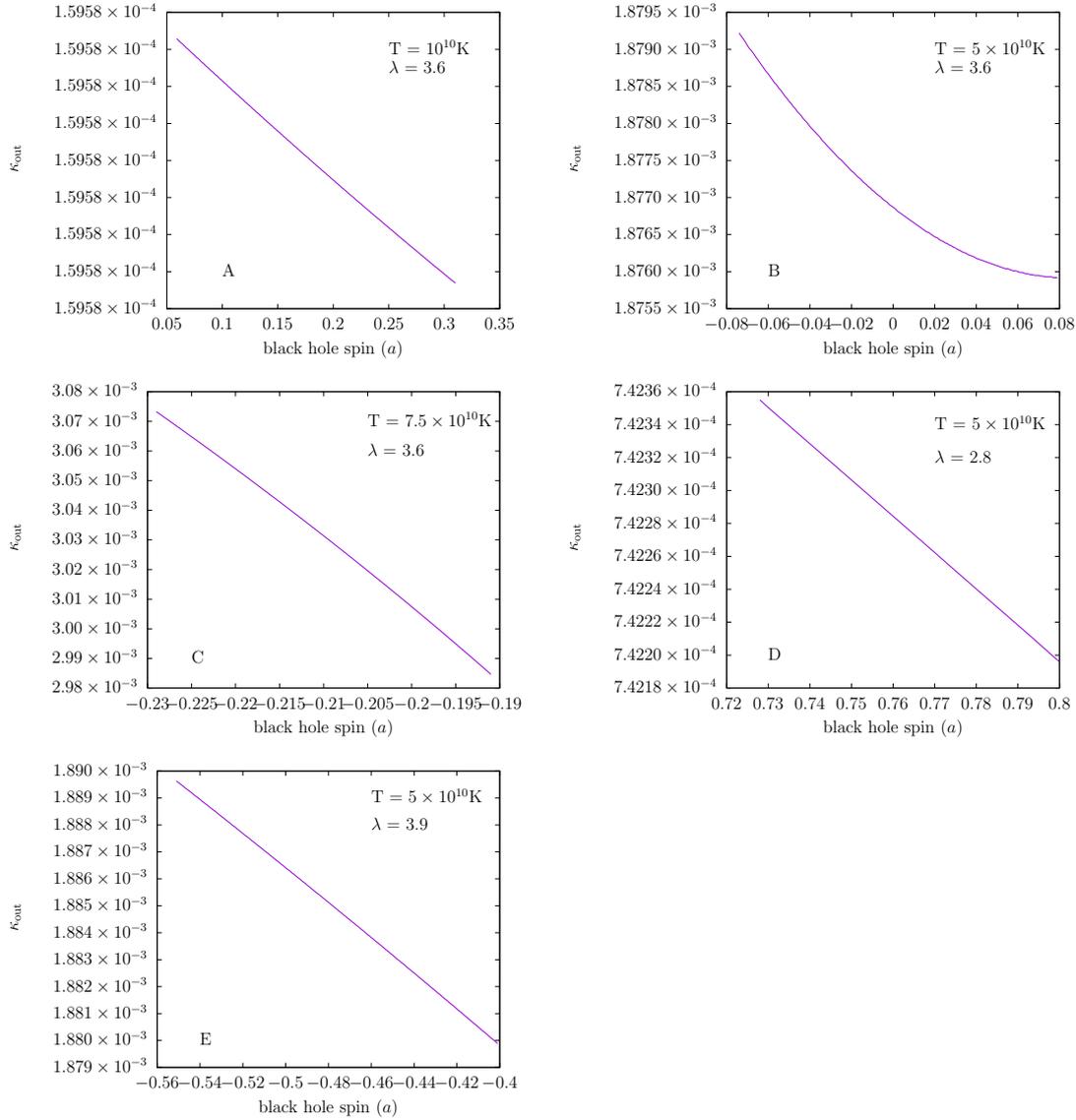

	\centering
	\begin{tabular}{cc}
		 \resizebox{0.45\textwidth}{!}{\input{surface_g_l36_T10_out}}&
		 \resizebox{0.45\textwidth}{!}{\input{surface_g_l36_T105_out}}\\
		 \resizebox{0.45\textwidth}{!}{\input{surface_g_l36_T1075_out}}&
		 \resizebox{0.45\textwidth}{!}{\input{surface_g_l28_T105_out}}\\
		 \resizebox{0.45\textwidth}{!}{\input{surface_g_l39_T105_out}}
	\end{tabular}
	\caption{Acoustic surface gravity at the outer horizon ($ \kappa_{\rm out} $) vs black hole spin ($ a $) plot. \textbf{A, B} and \textbf{C} corresponds to a fixed $ \lambda=3.6 $ and $ T = 10^{10}K,5\times10^{10}K $ and $ 7.5\times 10^{10}K $ respectively. \textbf{D, B} and \textbf{E} corresponds to a fixed $ T = 5\times10^{10}K $ and $ \lambda=2.8,3.6 $ and $ 3.9 $ respectively. For all the set of parameters $ [T,\lambda] $, $ \kappa_{\rm out} $ decreases with increasing $ a $ for prograde motion and increases with increasing $ a $ for retrograde motion.}\label{Fig:surface_g_outer}
\end{figure}

The acoustic metric given by Eq. (\ref{acoustic_metric}) is independent of time $t$. Therefore we have the stationary Killing vector $\chi^\mu=\delta^\mu_t$. The norm of the Killing vector $\chi^\mu$ on the  the acoustic horizon is given by $\chi^\mu\chi_\mu=G_{\mu\nu}\chi^\mu\chi^\nu=G_{tt}|_{\rm h}=0$. Hereafter in this section, by horizon we will mean acoustic horizon. Thus the stationary Killing vector is null on the horizon. The surface gravity in terms of the Killing vector $\chi^\mu$ which is null on the horizon is given by the relation as described in \cite{Poisson2004relativist,Wald,Bilic1999}
\begin{equation}\label{killing_eq}
\nabla_\alpha(-\chi^\mu\chi_\mu)=2\kappa\chi_\alpha
\end{equation}
which is to be evaluated on the horizon. Here $\kappa$ is the acoustic surface gravity. Now $\chi_\mu=G_{\mu\nu}\chi^\nu=G_{\mu\nu}\delta^\nu_t=G_{\mu t}$. Therefore from the $\alpha=r$ component of the Eq. (\ref{killing_eq}) the acoustic surface gravity is obtained to be
\begin{equation}
\kappa=\frac{1}{2G_{rt}}\partial_r(-G_{tt})|_{u_0^2=c_s^2}
\end{equation}
Hence
\begin{equation}\label{kappa1}
\kappa=\left|\frac{\sqrt{(g_{tt}g_{\phi\phi}+g_{\phi t}^2)(g_{\phi\phi}+2\lambda g_{\phi t}-\lambda^2 g_{tt})}}{(1-c_s^2)(g_{\phi\phi}+\lambda g_{\phi t})\sqrt{g_{rr}}}\frac{du_0}{dr}\right|_{\rm h}
\end{equation}
where the subscript ``h", as mentioned earlier, denotes that the quantities have been evaluated at the acoustic horizon. On the equatorial plane ($\theta=\frac{\pi}{2}$) the metric elements are 
\begin{equation}
g_{tt}=1-\frac{2}{r},\quad g_{\phi t}=-\frac{2a}{r},\quad g_{\phi\phi}=\frac{r^3+a^2r+2a^2}{r}
\end{equation}
Thus $\kappa$ can be written as
\begin{equation}\label{kappa2}
\kappa=\left|\frac{(r^2-2r+a^2)\sqrt{(r^3+a^2r+2a^2-4a\lambda-\lambda^2 (r-2))}}{(1-c_s^2)\sqrt{r}(r^3+a^2r+2a^2-2a\lambda)}\frac{du_0}{dr}\right|_{\rm h}
\end{equation}
In the Schwarzschild limit $a=0$, this reduces to the result derived earlier in \cite{arif:iso_sch}. Thus we obtain the acoustic surface gravity as a function of the background metric elements and the stationary values of the accretion variables. The surface gravity depends explicitly on the black hole spin $a$ and the specific angular momentum $ \lambda $. The acoustic surface gravity is linearly proportional to the gradient of the advective velocity ($ \frac{du_0}{dr} $) at the acoustic horizon. $ \frac{du_0}{dr} $ depends on the values of the parameters $ [T,\lambda,a] $, which could be found from numerical solution of the accretion flow. Thus the dependence of the acoustic surface gravity on the black hole spin $ a $ and the specific angular momentum $ \lambda $ could be understood only through numerical analysis of the accretion flow for a given set of  values of the  parameters $ [T,\lambda,a] $. In Fig. \ref{Fig:surface_g_inner} we plot the acoustic surface gravity at the inner acoustic horizon ($ \kappa_{\rm in} $) as a function of the black hole spin for a given set of values of $ [T,\lambda] $ and in Fig. \ref{Fig:surface_g_outer} we do the same for the acoustic surface gravity at the outer acoustic horizon ($ \kappa_{\rm out} $).  It is noticed that $ \kappa_{\rm in} $ increases with increasing $ a $ for prograde motion and decreases with increasing $ a $ for retrograde motion for a given $ [T,\lambda] $. Whereas $ \kappa_{\rm out} $ decreases with increasing $ a $ for prograde motion and increases with increasing $ a $ for retrograde motion for a given $ [T,\lambda] $. However since the shock forms only for a restricted region of parameter space the nature of the complete dependence of $ \kappa $ on $ a $ is difficult to understand explicitly. One can also notice that the value of $ \kappa_{\rm out} $ is up to $ 10^4 $ times smaller than that of $ \kappa_{\rm in} $. One of the reasons for this is that the gradient $ \frac{du_0}{dr} $ is smaller at the outer acoustic horizon than that at the inner acoustic horizon. Also since the outer acoustic horizon forms at a relatively large distance from the black hole, the Kerr parameter does not play any significant role to influence the properties of the spacetime close to the outer horizon. Therefore the value of the acoustic surface gravity  evaluated at the outer acoustic horizon does not seems to be reasonably sensitive on the black hole spin.

\section{Stability analysis}
We have derived the wave equation describing the propagation of the linear perturbation of velocity potential, mass accretion rate and the relativistic Bernoulli's constant in Eq. (\ref{w_velo_final}),(\ref{w_mass_final}) and (\ref{w_ber_final}) respectively. These equations can be further studied to understand the stability of the stationary accretion solutions. As the wave equations have similar forms, let us study one case, say for the velocity potential, and use the results accordingly for other cases. 

Let us take the trial solution as 
\begin{equation}
 \psi_1(r,t) = P_\omega (r) \e^{i\omega t} ,
 \end{equation} 
 using this trial solution in the wave equation $\partial_\mu(f^{\mu\nu}\partial_\nu \psi_1)=0$, where $f^{\mu\nu}$ is equal to $f_1^{\mu\nu}$ given by Eq. (\ref{f_velocity}), provides
\begin{equation}\label{Stability_general}
 -\omega^2 f^{tt}P_\omega+i\omega[f^{tr}\partial_rP_\omega+\partial_r(f^{rt}P_\omega)]+\partial_r(f^{rr}\partial_rP_\omega)
 =0
 \end{equation} 

\subsection{Standing wave analysis}
For standing wave to form there must be two nodes, one at some inner point $r_1$ and another at some outer point $r_2$, such that at these two points we have $P_\omega(r_1)=0=P_\omega(r_2)$. In other words the perturbation must vanish for all times at two different radii. Multiplying the Eq. (\ref{Stability_general}) by $P_\omega(r)$ and integrating the resulting equation between $r_1$ and $r_2$ gives
\begin{equation}\label{multiplied_by_p}
\omega^2 \int_{r_1}^{r_2} P_\omega^2 f^{tt}dr-i\omega\int_{r_1}^{r_2}\partial_r[f^{tr}P_\omega^2]dr-\int_{r_1}^{r_2}[P_\omega \partial_r(f^{rr}\partial_r P_\omega)]dr=0
\end{equation}
The middle term in the above equation does not contribute as at the boundary $r_1$ and $r_2$, $P_\omega$ vanishes. Integrating the last term by parts, the Eq. (\ref{multiplied_by_p}) can be written as
\begin{equation}
\omega^2 \int_{r_1}^{r_2} P_\omega^2 f^{tt}dr+\int_{r_1}^{r_2} f^{rr}(\partial_r P_\omega)^2dr=0,
\end{equation}
and thus we get 
\begin{equation}\label{omega_for_standing}
\omega^2 = -\frac{\int_{r_1}^{r_2} f^{rr}(\partial_r P_\omega)^2dr}{\int_{r_1}^{r_2} f^{tt} P_\omega^2 dr}
\end{equation}
One thing to be noticed is that the inner boundary condition $P_\omega(r_1)=0$ may be satisfied only if the accretor has a physical surface. In that case, the outer boundary could be located at the source from which the accreting material is coming and the inner boundary could be located at the surface of the accretor which is accreting the material (see Petterson \etal \cite{Petterson1980}). Also, the flow should be continuous in the whole range in between these two boundary points. If the  accretor is a neutron star then the surface (where the inner boundary with vanishing perturbation is to be located) should be separated from a possible supersonic region by a shock formation. This would imply that the solution, in that case, would not be continuous in the range between the outer boundary point and the surface of the neutron star and therefore standing wave analysis could not be performed. Also in the black hole accretion, the flow enters the horizon supersonically (\cite{Frank1985accretion,Liang1980transonic}) and there is no mechanism to make the perturbation vanish and therefore the above-mentioned requirements are not expected to be fulfilled and hence standing wave may not be formed in the context of black hole accretion.  Also, the flow has to be subsonic for the whole range as transonic flow would contain a horizon from which the reflected wave cannot come out and superpose with the wave in the outer region. Therefore the standing wave analysis, which relies on the continuity of the solution, is restricted to only subsonic flows.  However, provided that the flow is subsonic in a particular astrophysical system (e.g., in case of accretion onto a Newtonian star, depending on the location of the surface of the star, the accretion flow may be subsonic for all radial distance and hit the surface of the star subsonically \cite{CHAKRABARTI_physics_Reports}) we can study Eq. (\ref{omega_for_standing}) to understand the nature of $\omega$. The stability analysis for accretion onto a compact object in flat spacetime was done by Petterson \etal \cite{Petterson1980} where the flow was considered to be subsonic.

From Eq. (\ref{f_velocity}) $f^{tt}$ for velocity potential is given by
\begin{equation}
f^{tt} = -\frac{\sqrt{-\tilde{g}}H_0}{\rho_0^{c_s^2-1}}[-g^{tt}+(v^t_0)^2(1-\frac{1}{c_s^2})]
\end{equation}
as $g^{tt}> 0$ and $c_s^2<1$ we find that $f^{tt}>0$. Now $f^{rr}$ is given by
\begin{equation}
f^{rr} = -\frac{\sqrt{-\tilde{g}}H_0}{\rho_0^{c_s^2-1}}[g^{rr}+(v^r_0)^2(1-\frac{1}{c_s^2})].
\end{equation}
Using Eq. (\ref{v_in_CRF}) the terms inside the square bracket can be written as
\begin{equation}
g^{rr}+\frac{u_0^2}{g_{rr}(1-u_0^2)}(1-\frac{1}{c_s^2}) = \frac{(1-u_0^2)+u_0^2(1-\frac{1}{c_s^2})}{g_{rr}(1-u_0^2)}=\frac{\left(1-\frac{u_0^2}{c_s^2}\right)}{g_{rr}(1-u_0^2)}>0
\end{equation}
Thus $f^{rr}<0$ and hence $\omega^2>0$. Therefore $\omega$ has two real roots and the trial solution is oscillatory and the stationary accretion solution is stable. We have used the fact that the flow is subsonic to get $f^{rr}<0$. Same result is also applicable for the relativistic Bernoulli's constant. It is easy to show that the conclusion also holds for the mass accretion rate.

\subsection{Traveling wave analysis} 
Following Petterson \etal \cite{Petterson1980} we study the traveling waves whose wavelengths are  small compared to the smallest length scale in the system. In case of black hole accretion,  this may be the radius of the event horizon of the black hole. Therefore for such wave the frequency is large and hence the trial solution may be taken as the power series of the form
\begin{equation}\label{series_solution}
P_\omega(r)= \exp \left[\sum_{n=-1}^{\infty} \frac{k_n(r)}{\omega^n} \right]
\end{equation}
We substitute the trail solution in Eq. (\ref{Stability_general}) and find out leading order terms by equating the coefficients of individual power of $\omega$ to zero. Thus we get
\begin{eqnarray}\label{omega2}
& \textrm{coefficient of } \omega^2:  f^{rr}(\partial_rk_{-1})^2 + 2i f^{tr}\partial_r k_{-1} -f^{tt}= 0\\ 
& \textrm{coefficient of } \omega: f^{rr}\left[\partial_r^2k_{-1}+2\partial_rk_{-1}k_0\right]+i[2f^{tr}\partial_r k_0 \nonumber \label{omega}\\ 
& +\partial_r f^{tr}]+\partial_rf^{rr}\partial_r k_{-1} = 0\\ 
& \textrm{coefficient of } \omega^0: f^{rr}[\partial_r^2 k_0 + 2\partial_r k_{-1}\partial_r k_1+(\partial_r k_0)^2]\nonumber \\
& +\partial_r f^{rr}\partial_r k_0+2i f^{tr}\partial_r k_1=0 \label{omega0}
\end{eqnarray}
Eq. (\ref{omega2}) gives 
\begin{equation}\label{k-1}
k_{-1}(r) = i \int \frac{-f^{tr}\pm \sqrt{(f^{tr})^2-f^{tt}f^{rr}}}{f^{rr}}dr
\end{equation}
using $k_{-1}(r)$ from Eq. (\ref{k-1}) in Eq. (\ref{omega}) gives 
\begin{equation}
\label{k0}
k_0(r) = -\frac{1}{2} \ln [\sqrt{(f^{tr})^2-f^{tt}f^{rr}}] + \textrm{constatnt}
\end{equation}
and using Eq. (\ref{k-1}) and (\ref{k0}) in Eq. (\ref{omega0}) gives 
\begin{equation}
\label{k1}
k_1(r) =\pm \frac{i}{2} \int \frac{\partial_r(f^{rr}\partial_r k_0)+f^{rr}(\partial_r k_0)^2}{\sqrt{(f^{tr})^2-f^{tt}f^{rr}}}dr 
\end{equation}
Now for the case velocity potential or relativistic Bernoulli's constant
\begin{equation}
\textrm{det} f^{\mu\nu} = f^{tt}f^{rr}-(f^{rt})^2 =\left( \frac{\sqrt{-\tilde{g}}H_0}{\rho_0^{c_s^2-1}}\right)^2\mathcal{F}
\end{equation}
and for the case of mass accretion rate
\begin{equation}
\textrm{det} f^{\mu\nu} = f^{tt}f^{rr}-(f^{rt})^2 =\left( \frac{g_{rr}v_0^r c_s^2}{v^t_0 v_{t0}\tilde{\Lambda}}\right)^2\mathcal{F}
\end{equation}
where $v_{t0}$ is the stationary value of $v_t$ given by Eq. (\ref{v_t}) and $v^r_0$ and $v^t_0$ are stationary values of $v^r$ and $v^t$ given by Eq. (\ref{v_in_CRF}) and (\ref{vt_0}) respectively and
\begin{equation}\label{stability_F}
\mathcal{F}=[-g^{tt}g^{rr}+(1-\frac{1}{c_s^2})(-g^{tt}(v^r_0)^2+g^{rr}(v^t_0)^2)].
\end{equation}
We can further express $\mathcal{F}$ in terms of $\lambda,u_0$ and the background metric elements as
\begin{equation}
\mathcal{F} = -\frac{g_{\phi\phi}}{g_{rr}(g_{\phi\phi}g_{tt}+g_{\phi t}^2)}\left[1+\frac{(1-c_s^2)}{c_s^2(1-u_0^2)}\left(\frac{(1+\lambda \frac{g_{\phi t}}{g_{\phi \phi}})^2}{(1+2\lambda \frac{g_{\phi t}}{g_{\phi\phi}}-\lambda^2 \frac{g_{tt}}{g_{\phi\phi}})} -u_0^2\right) \right]
\end{equation}
$\mathcal{F}$ is negative everywhere. This can be understood in the following way: The expression of $v^t$ from Eq. (\ref{vt_0}) requires $g_{\phi\phi}+2\lambda g_{\phi t}-\lambda^2 g_{tt}>0$ in order that $v^t$ is real. Which can be rewritten as $\left(1+\lambda \frac{g_{\phi t}}{g_{\phi\phi}}\right)^2-\lambda^2\left(\frac{g_{\phi t}^2}{g_{\phi \phi}^2}+\frac{g_{tt}}{g_{\phi\phi}}\right)>0$ or 
\begin{equation}
0<1-\lambda^2 \left[{(\frac{g_{\phi t}^2}{g_{\phi \phi}^2}+\frac{g_{tt}}{g_{\phi\phi}})}/{(1+\lambda \frac{g_{\phi t}}{g_{\phi \phi}})^2}\right]<1
\end{equation}
 Therefore
\begin{equation}
\frac{(1+\lambda \frac{g_{\phi t}}{g_{\phi \phi}})^2}{(1+2\lambda \frac{g_{\phi t}}{g_{\phi\phi}}-\lambda^2 \frac{g_{tt}}{g_{\phi\phi}})} = \frac{1}{1-\lambda^2 \left[{(\frac{g_{\phi t}^2}{g_{\phi \phi}^2}+\frac{g_{tt}}{g_{\phi\phi}})}/{(1+\lambda \frac{g_{\phi t}}{g_{\phi \phi}})^2}\right]}>1
\end{equation}
using the fact that $u_0^2<1$ and $c_s^2<1$ it is easy to see that $\mathcal{F}$ is negative everywhere. Thus $k_{-1}(r)$ and $k_1(r)$ are purely imaginary. Therefore the leading contribution to the amplitude of the wave comes from $k_0(r)$. Thus considering the first three terms in the expansion in Eq. (\ref{series_solution}) the amplitude of the wave can be approximated as
\begin{equation}
\label{wave_amplitude}
|\psi_1|=|\xi_1|\approx  \left[\frac{\rho_0^{2c_s^2-2}}{\tilde{g}H^2_0\mathcal{F}} \right]^{\frac{1}{4}},\quad |\Psi_1|\approx \left[\left(\frac{v^t_0v_{t0}\tilde{\Lambda}}{g_{rr}v^r_0 c_s^2}\right)^2 \frac{1}{-\mathcal{F}}\right]^{\frac{1}{4}}
\end{equation}
The trial solution in Eq. (\ref{series_solution}) with the frequency $\omega\gg 1$ ensures that contribution from the higher order terms will be very small. The amplitude given by Eq. (\ref{wave_amplitude}) is bounded and the solution is therefore stable.

\section{Concluding remarks}
By constructing the phase portraits of the stationary accretion solutions and the corresponding causal structures we were able to show that the critical points of the accretion flow are indeed the horizons in the acoustic spacetime. The causal structure with shock also shows that the location of the shock formation corresponds to the presence of an acoustic white hole. The main goal of Unruh's work \cite{Unruh} was to show that a transonic fluid system shows Hawking radiation like effect and the quantized phonons emitted from the acoustic horizon has a thermal spectrum with an analogue Hawking temperature $ T_{AH} $. The Hawking temperature (as measured at infinity) of a black hole is given in terms of the surface gravity $ \kappa_g $\footnote{The subscript $ g $ is used to denote black hole surface gravity as we have already denoted the acoustic surface gravity by $ \kappa $.} as $ T_H = \frac{\hbar\kappa_g}{2\pi k_B}$ (in the units we are working with)\cite{Hawking1974,Hawking1975}. The analogue Hawking temperature could be given by a similar formula with the acoustic surface gravity $ \kappa $ and hence the acoustic surface gravity provides a way to evaluate the corresponding analogue Hawking temperature $ T_{AH} $. By plotting the location of the acoustic horizon and the corresponding acoustic surface gravity as a function of the black hole spin parameter $ a $ we were able to show how the background metric influences the acoustic spacetime properties, which was one of the main goals of the present work. However, we have not discussed how the static limit surface may influence the acoustic specetime properties in the present work and such analysis may be reported elsewhere.

In the present work, we have considered only one model of accretion disc in which the flow geometry of the accretion disc is considered to be conical in shape such that the local thickness $ H(r) $ of the flow at any radial distance is given by $ H(r)\propto r $. However, there exist other models of accretion also such as constant height flow (CH) and flow with equilibrium along the vertical direction (VE)\cite{Bilic-Choudhury}.
In CH model the flow thickness is assumed to be constant at all radial distances, i.e, $ H(r) = {\rm constant} $. Whereas in VE model the flow is considered to be in hydrostatic equilibrium along the vertical direction and for such model the local flow thickness is given by the relation \cite{Abramowicz1996ap}
\begin{equation}
-\frac{2p}{\rho}+\frac{H^2}{r^4}(v_\phi^2-a^2(v_t-1)) = 0.
\end{equation}
Thus it could be noticed that for the conical and CH models the flow  thickness does not depend on any accretion flow variables such the velocities, pressure or density of the flow, whereas in case of VE model the flow thickness depends on the accretion flow variables as well as the black hole spin in a rather complicated way. While performing the linear perturbation analysis in the conical model the flow thickness remains unperturbed as it is independent of the flow variables which are perturbed linearly (See section \ref{Sec:Linear_pert}). Due to the same reason in CH model also the flow thickness would remain unperturbed. Therefore one could expect the linear perturbation analysis to be identical to that for the conical model presented in section \ref{Sec:Linear_pert} and hence the form of the acoustic metric would be also the same. (The actual values of different quantities, such as the location of the acoustic horizon, causal structure or the acoustic surface gravity, for CH model though will be different  from that of conical model due to the difference in flow geometries which give rise to different equations for gradient of advective velocity as well as different constants for the integral of the continuity equation.) However in case of VE model, due to the dependence of the flow thickness on the flow variables, the flow thickness also gets perturbed. Therefore the acoustic spacetime metric for accretion in VE model is expected to be different than that for the conical flow. Such analysis would be too complicated to be presented here and may be reported elsewhere.

We have considered inviscid accretion of ideal fluid only. Description of non ideal fluid is characterized by various dissipative processes in conjunction with the presence of viscosity. Viscosity, however, breaks the Lorentzian symmetry \cite{Barcelo} and acoustic metric cannot be constructed for viscous flow using the formalism we follow in this work. It should also be mentioned that effect of viscosity, as well as magnetic field, may not always be neglected as the dissipative mechanism, through different processes such as comptonisation, bremsstrahlung or synchrotron processes, may become significant. This would influence the overall flow dynamics and therefore turbulent instability may develop. In such cases the linear stability analysis would also become insufficient and non-linear stability analysis would be required to ensure the stability of the system. Such work would require large scale numerical simulation of the flow profile and analysis for such flow is beyond the scope of the present work. Hence the study of the viscous accretion of non ideal fluid as well as analogue system is beyond the scope of this work. We have also assumed that the accretion flow is axially symmetric. Thus for any non-axially asymmetric flow, the present formalism would not be appropriate.

For large scale astrophysical fluid flows, transient phenomena are not quite uncommon to take place. For accreting black hole systems, any conclusion drawn based on the results obtained using the integral stationary transonic solutions are thus reliable only if the accretion flow under considerations happens to be steady. One thus needs to ensure whether such steady flow is stable, at least within a reasonable astrophysical time scale. Such cross verification can be accomplished by perturbing the corresponding spacetime dependent governing equations (the Euler and the continuity equations for the present case) governing the flow and by investigating whether such perturbation converges (or, at least does not diverge) to ensure the stability of such transonic accretion. In the present work, a linear perturbation scheme has been followed to obtain the emergent relativistic acoustic geometry embedded within the accretion flow, as well as to check whether the background steady state of such flow is stable. This also implies that the emergent gravity phenomena are a natural outcome of the linear stability analysis of transonic accretion. Our present work, thus, 
bridges a gap between two apparently different topics -- the stability analysis of astrophysical accretion as well as the classical analogue gravity phenomena -- for general relativistic isothermal accretion onto rotating black holes. To the best of our knowledge, 
such works were not available in the literature before. 

\section{Acknowledgment}
The author would like to thank his PhD supervisor Tapas K. Das for introducing him to the problem and for helping him in preparing the manuscript, and Pratik Tarafdar for useful discussions. The author would also like to thank the anonymous referees for providing useful comments and suggestions.

\section*{References}
\bibliography{reference_arif} 
\end{document}